\documentclass[useAMS,usenatbib]{mn2e}
\usepackage{amsmath,epsf}
\def\lesssim{\mathrel{\hbox{\rlap{\hbox{\lower4pt\hbox{$\sim$}}}\hbox{$<$}}}}
\def\gtrsim{\mathrel{\hbox{\rlap{\hbox{\lower4pt\hbox{$\sim$}}}\hbox{$>$}}}}
\newcommand{\ltaraw}{$\; \buildrel < \over \sim \;$}
\newcommand{\lta}{\lower.5ex\hbox{\ltaraw}}
\newcommand{\gtaraw}{$\; \buildrel > \over \sim \;$}
\newcommand{\gta}{\lower.5ex\hbox{\gtaraw}}

\newcommand{\ie}{{\it i.e.~}}
\newcommand{\eg}{{\it e.g.~}}

\newcommand{\etal}{{\it et.~al.~}}

\newcommand{\Chandra}{{\it Chandra}}
\newcommand{\XMM}{{\bf XMM}}

\def\apj{ApJ}%
\def\apjl{ApJ}%
\def\aap{A\&A}%
\def\mnras{MNRAS}%
\newcommand{\LT}{{$L_{x}-T_{x}$}}
\newcommand{\ML}{{$M_{t}-L_{x}$}}
\newcommand{\MT}{{$M_{t}-T_{x}$}}

\newcommand{\trelax}{$t_{relax}$}

\newcommand{\taccrete}{$t_{accrete}$}
\newcommand{\tclosest}{$t_{closest}$}
\newcommand{\tapo}{$t_{apo}$}

\newcommand{\kpc}{\,\mbox{kpc}}

\newcommand{\Mpc}{\,\mbox{Mpc}}

\newcommand{\eV}{\,\mbox{eV}}

\newcommand{\Gyr}{\,\mbox{Gyr}}

\newcommand{\keVcmsq}{\,\mbox{keVcm$^{2}$}}
\newcommand{\ks}{\,\mbox{ks}}

\newcommand{\keV}{\,\mbox{keV}}
\newcommand{\msun}{\,$M_{\odot}$}

\newcommand{\Gyrs}{\,\mbox{Gyrs}}
\newcommand{\Myrs}{\,\mbox{Myrs}}

\title[The impact of mergers on relaxed X-ray clusters II]
  {The impact of mergers on relaxed X-ray clusters \\
   II. Effects on global X-ray and SZ properties and their scaling relations}
\author[Gregory B. Poole \etal]{Gregory B. Poole$^{1}$\thanks{E-mail: gbpoole@uvic.ca}, Arif Babul$^1$, Ian G. McCarthy$^2$, Mark A. Fardal$^3$, \newauthor
   C. J. Bildfell$^{1}$, Thomas Quinn$^4$ and Andisheh Mahdavi$^{1}$\\
  $^{1}$Dept. of Physics \& Astronomy, University of Victoria,
  Elliott Building, 3800 Finnerty Rd., Victoria, BC, V8P 1A1, Canada\\
  $^2$Department of Physics, University of Durham,
      South Road, Durham DH1 3LE, UK\\
  $^3$Dept.\ of Astronomy, University of Massachusetts,
      Amherst, MA, 01003, USA\\
  $^4$Dept.\ of Astronomy, University of Washington, Seattle, WA 98195
  }

\date{draft version \today}
\pagerange{\pageref{firstpage}--\pageref{lastpage}} \pubyear{2006}

\def\LaTeX{L\kern-.36em\raise.3ex\hbox{a}\kern-.15em
    T\kern-.1667em\lower.7ex\hbox{E}\kern-.125emX}

\begin{document}

\label{firstpage}

\maketitle
\begin{abstract}
We use the suite of simulations presented in \citet{P06a} to examine global X-ray and Sunyaev-Zel'dovich (SZ) observables for systems of merging relaxed X-ray clusters.  The time evolution of our merging systems' X-ray luminosities, temperatures, total mass measures, SZ central Compton parameters and integrated SZ fluxes are presented and the resulting impact on their scaling relations examined.  In all cases, and for all parameters, we observe a common time evolution: two rapid transient increases during first and second pericentric passage, with interceding values near or below their initial levels.  This is in good qualitative agreement with previous idealized merger simulations \citep[\eg][]{RickerandSarazin01}, although we find several important differences related to the inclusion of radiative cooling in our simulations.  These trends translate into a generic evolution in the scaling-relation planes as well: a rapid transient roughly along the mass scaling relations, a subsequent slow drift across the scatter until virialization, followed by a slow evolution along and up the mass scaling relations as cooling recovers in the cluster cores.  However, this drift is not sufficient to account for the observed scatter in the scaling relations.  We also study the effects of mergers on several theoretical temperature measures of the intracluster medium: emission weighted measures ($T_{ew}$), the spectroscopic-like measure proposed by \citet[][$T_{sl}$]{Mazzottaetal04} and plasma model fits to the integrated spectrum of the system ($T_{spec}$).  We find that $T_{sl}$ tracks $T_{spec}$ for the entire duration of our mergers, illustrating that it remains a good tool for observational comparison even for highly disturbed systems.  Furthermore, the transient temperature increases produced during first and second pericentric passage are $15-40$\% larger for $T_{ew}$ than for $T_{sl}$ or $T_{spec}$.  This suggests that the effects of transient temperature increases on $\sigma_8$ and $\Omega_M$ derived by \citet{Randalletal02} are over estimated.  Lastly, we examine the X-ray SZ proxy proposed by \citet{Kravtsovetal06} and find that the tight mass scaling relation they predict remains secure through the entire duration of a merger event, independent of projection effects.

\end{abstract}
\section{Introduction}\label{sec-intro} 
It has been known for some time that clusters obey simple and relatively well defined power-law scalings between many of their globally integrated or averaged properties (\eg\ temperature, X-ray luminosity and mass).  Due to their proven utility in cosmological studies \citep[see][for an excellent review]{Voit05}, their use as constraints on theoretical models of galaxy formation \citep{BBLP} and the significant discrepancy between observations and simple early theoretical expectations \citep[\eg~][]{Kaiser86,Markevitch98}, these relations have received a great deal of scrutiny in recent years.  

Significant progress has recently been made towards reconciling theoretical expectations for the slope and normalization of these relations with observations \citep[\eg~][]{Borganietal04,Voitetal02}.  As a result, attention is now turning towards accounting for their observed scatter.  One significant early step towards this goal was made by \citet[][M04 hereafter]{MBBPH} who illustrated that the position of systems relative to the mean observed \LT~ and \ML~ relations correlates with the system's central entropy, and hence the morphology of a system's core.  Specifically, they found that systems with compact cool cores (CCCs; classically identified as ``cooling flow'' systems) and low central entropies tend to lie on the high-luminosity/low-temperature sides of observed distributions while systems lacking compact cool cores (NCCs; classically identified as ``non-cooling flow'' systems) and possessing elevated central entropies lie on the low-luminosity/low-mass sides.  Recently, \citet{OHaraetal05} have demonstrated that these two populations are indistinguishable in these planes when their cores are removed during analysis.  This suggests that variations in core properties are primarily responsible for the scatter in observed scaling relations, supporting previous findings of \citet{Fabianetal94} and \citet{Markevitch98}.

These results are corroborated by \citet{Baloghetal06} who find that variations in dark matter structure and uncertainties in cosmological models can not account for the scatter in the $L_x-T_x$ or $M_t-L_x$ scaling relations.  They suggest that variations in the minimum level of core entropy due to differences in heating and/or cooling efficiencies of core material are responsible.  Little speculation is made regarding the physical origins of such efficiency variations but they theorize that mergers are not likely the source, citing qualitative observations made by authors of previous theoretical merger studies that mergers generally drive systems to evolve parallel to observed scaling relations \citep{RTK04}.  This lies in stark contrast to the work of \citet{Smithetal05} however, who use strong lensing measurements and high resolution \Chandra~ observations to illustrate a correlation between the degree to which local cluster cores are disturbed and their variance from the mean \ML, \MT, and \LT~ scaling relations.  Hence, several unresolved issues regarding the role of mergers in shaping the scatter in observed cluster scaling relations persist.  

Scaling relations involving the Sunyuev-Zeldovich (SZ) properties of clusters, although previously considered \citep{McCarthyetal03,McCarthyetal03b,CavaliereMenci01}, have received significantly less attention than those involving mass, X-ray luminosity and temperature.  With the impending availability of large and statistically significant datasets of SZ observations, a careful examination of the effects of mergers on SZ scaling relations is warranted.  In this paper we will use a set of nine idealized two-body cluster merger simulations \citep[first presented in][Poole06a henceforth]{P06a} to examine the effects of mergers on cluster scaling relations \citep[refer to Poole06a for a review of pioneering work involving such simulations and][RS01 hereafter, for a particularly good first examination of some of the issues revisited in this work]{RickerandSarazin01}.  Our simulations have been constructed to probe a representative range of impact parameters and mass ratios of mergers between CCC systems with a primary (most massive system) of $10^{15}$\msun .  A detailed description of these simulations can be found in Poole06a. 

In Section \ref{sec-method} we will briefly summarize several aspects of our method for constructing cluster mergers presented in Poole06a which are relevant to this paper.  In Section \ref{analysis-LT} we present the temporal evolution and scaling relation of our systems' X-ray luminosities ($L_x$) and temperatures ($T_x$), examining the effects of various temperature measures and of core excision on the behaviour of these quantities.  In Section \ref{analysis-mass} we examine the effectiveness of hydrostatic estimates of total mass ($M_t$), its isothermal $\beta$-model implementation and the \MT~ and \ML~ scaling relations which are obtained from them. In Section \ref{analysis-SZ} we study the evolution of our systems' SZ properties (both central Compton parameters and integrated SZ fluxes), examining the effectiveness of the $\beta$-model approximation often implemented by observers as well as the most relevant scaling relations which can be constructed from them.  In Section \ref{analysis-XSZ} we examine the effects of mergers on the X-ray SZ proxy recently proposed by \citet{Kravtsovetal06}.  In Section \ref{sec-discussion} we discuss several issues raised by our work and finally, we summarize the most notable results of our study in Section \ref{sec-summary}.

In all cases our assumed cosmology will be ($\Omega_M$,$\Omega_\Lambda$)=($0.3$,$0.7$) with $H_o$=$75$km/s/Mpc.

\section{Simulations}\label{sec-method}

We run our simulations with GASOLINE \citep{wadsley04}, a versatile parallel SPH tree-code with multi-stepping.  We include the effects of radiative cooling in our simulations and feedback from star formation but the effects of jets from active galactic nuclei (AGN) are omitted.  

The basis of our study is a set of 9 idealized 2-body cluster merger simulations.  Three mass ratios are examined (1:1, 3:1 and 10:1) with the most massive system set to have a mass of $10^{15}$\msun~ in every case.  For each mass ratio, three impact parameters are examined (one head-on and two off-axis cases).  We express these impact parameters with the quantity $v_t/V_c$, where $v_t$ is the relative tangential velocity of the interacting systems when the core of the smaller system crosses the virial radius of the larger system and $V_c$ is the larger system's circular velocity at its virial radius.

In Poole06a we present a detailed account of how our isolated clusters and their orbits are initialized as well as the numerical methods and code parameters of our simulations.  In this section we summarize key aspects of our approach pertinent to our present analysis and direct readers looking for additional details to Poole06a.

\subsection{Initial conditions}\label{numerics-cluster_initial}

\begin{table*}
\begin{minipage}{175mm}
\caption[Deviations of remnant global properties from expected mass scaling relations]{Percentage difference of the global properties of our systems from the values required to preserve accepted mass scaling relations at the end of our simulations and at \trelax~ (in parenthesis).  For X-ray observables, observed mass scaling relations are used while for SZ observables ($y_o$ and $S_\nu$), mass scaling relations computed from our fiducial theoretical models are assumed (see Section \ref{analysis-L_T}).  For temperature we use the mass scaling relation of \citet{Horner_thesis} and for luminosity we use the mass scaling relation of \citet{ReiprichBohringer02}  Corrected results are for analysis conducted with the central $0.1R_{200}$ excised and uncorrected values are performed without excising the central regions. \label{table-scaling}}
\begin{tabular}{ccrlrlrlrlrlrl}
\hline
$M_p$:$M_s$                             &
$v_t/v_c$                               &
\multicolumn{2}{c}{$L_x$ (uncorrected)} & 
\multicolumn{2}{c}{$L_x$ (corrected)}   & 
\multicolumn{2}{c}{$T_x$ (uncorrected)} & 
\multicolumn{2}{c}{$T_x$ (corrected)}   & 
\multicolumn{2}{c}{$y_o$}               & 
\multicolumn{2}{c}{$S_\nu(R_{2500})$}  \\
\hline
1:1   &  0.00 &  21.1 & (-13.1) & -13.0 & (-2.0)  & 10.3 & (5.5)  &  2.9 & (-1.5) &  28.8 & (-3.3)  & 13.4 & (14.7) \\
1:1   &  0.15 & -25.6 & (-38.3) & -13.3 & (-0.6)  &  8.2 & (13.1) & -1.3 & (-0.4) &  -8.7 & (-27.6) & 10.0 & (20.8) \\
1:1   &  0.40 & -49.6 & (-52.9) & -27.6 & (-26.1) &  7.5 & (9.2)  & -3.1 & (-4.9) & -24.2 & (-39.3) &  3.9 & (6.7)  \\
3:1   &  0.00 & -13.3 & (-27.1) &   3.3 & (14.6)  & 10.9 & (14.7) &  2.1 & (2.2)  & -17.2 & (-37.7) & 12.9 & (24.7) \\
3:1   &  0.15 & -38.6 & (-40.4) &  -3.2 & (5.7)   & 14.6 & (13.7) &  0.9 & (0.3)  & -29.8 & (-47.6) & 11.8 & (18.2) \\
3:1   &  0.40 & -31.7 & (-35.2) & -24.9 & (-16.0) &  5.3 & (9.6)  & -1.7 & (0.0)  & -15.0 & (-33.6) & -6.6 & (1.6)  \\
10:1  &  0.00 &   5.7 & (12.3)  & -10.7 & (19.1)  &  3.7 & (13.1) & -1.0 & (4.4)  &  -6.2 & (-11.4) &  2.4 & (20.4) \\
10:1  &  0.15 & -16.0 & (-18.4) & -13.5 & (7.7)   &  6.5 & (12.7) &  0.6 & (3.9)  & -18.0 & (-23.8) & -2.6 & (12.0) \\
10:1  &  0.40 & -17.5 & (1.4)   & -23.3 & (-7.7)  &  5.8 & (11.8) & -0.2 & (4.4)  & -21.6 & (-7.6)  & -7.2 & (7.1)  \\
\hline
\end{tabular}
\end{minipage}
\end{table*}

We have chosen to focus our study on systems initialized to posses compact cores ($r_c \sim 50$\kpc).  We have chosen this initial configuration because it is the natural convergent state for a relaxed cluster in the absence of feedback and representative of the structure of a large fraction of clusters \citep{Peresetal98,Edgeetal92}.  In the context of hierarchical clustering, such systems are well understood: they are systems which have remained undisturbed long enough for vigorous radiative cooling to have produced a positive central temperature gradient and high central gas densities.

To initialize the structure of our systems, we have followed the analytic prescription of \citet{BBLP} and \citet{MBBPH} to produce clusters which conform with recent theoretical and observational insights into cluster structure.  The dark matter density profiles of our systems follow an NFW-like form \citep{NFW96,Mooreetal98} with the central asymptotic logarithmic slope chosen to be $\beta=1.4$ and the scale radius ($r_s$) selected to yield a concentration $c=R_{200}/r_s=2.6$ (we will use $R_{\Delta}$ throughout to indicate the radius within which the mean density of the system is $\Delta$ times the critical density, $\rho_c=3 H_0^2/8 \pi G$; $R_{200}=1785$\kpc, $R_{500}=1166$\kpc\ and $R_{cool}=180$\kpc\ initially for our primary systems).  The initial gas density and temperature profiles of the clusters are set by requiring that (1) the gas be in hydrostatic equilibrium within the halo, (2) the ratio of gas mass to dark matter mass within the virial radius be $\Omega_b/(\Omega_m - \Omega_b)$, and (3) the initial gas entropy$^5$\footnotetext[5]{We use the standard proxy for entropy given by $S\equiv kT/n_e^{2/3}$ with $n_e$ and $T$ representing the electron density and temperature of the gas.} scale as  $S(r) \propto r^{1.1}$ over the bulk of the cluster body.  We normalize the entropy profiles such that the temperature of the ICM at $R_{vir}$ is half the virial temperature.

We construct orbits for our systems which produce specified radial and tangential velocities for the secondary system ($v_r$ and $v_t$ respectively) when its centre of mass reaches the virial radius of the primary ($R_{vir}$).  For each of the three mass ratios we study, we examine three orbits selected to produce a typical value of $v_r(R_{vir})$ and to cover a significant range of the transverse velocity $v_t(R_{vir})$ giving rise to mergers found in cosmological dark matter simulations \citep{tormen97,vitvitska02}.  

Throughout our analysis we shall refer to three coordinate axes; $x$ and $y$ will denote directions in the plane of the initial orbit (with the initial separation of the systems being along the $x$ axis) and $z$ the direction orthogonal to this plane.  We shall also distinguish the two systems by calling the more massive system the ``primary'' system (we arbitrarily choose one to be the primary in the 1:1 cases, with no consequences for our results given their high degree of symmetry) and the less massive system as the ``secondary''.

\subsection{Dynamical evolution}\label{sec-dynamics}

In Poole06a we illustrated the generic dynamical progression which each of our simulations proceed through, identifying five distinct stages: a pre-interaction phase, first core-core interaction, apocentric passage, secondary core accretion, and relaxation.  These stages feature prominently in the discussion which follows so we will briefly review their progression here, for those who have not read Poole06a.

As the cores accelerate towards each other during pre-interaction, a pair of shock fronts materialize and are driven towards each core, heating and compressing them briefly.  At \tclosest~ the cores of the two systems reach closest approach and these effects reach their maximum strength.  In every case (including the head-on collisions), some part of the secondary's core survives its first encounter with the primary, tidally stripped into large cool streamers in off-axis cases and strings of several clumps in the on-axis cases.  At \tapo~ the disturbed cores reach maximum separation.  After subsequently reaching second pericentric passage at \taccrete, no observable trace of the secondary core remains.  This period marks the beginning of a prolonged period of accretion when the plume of material generated from the disruption of the secondary's core accretes as a high velocity stream.  This instigates a second episode of core heating followed by a period of relaxation.  At approximately \trelax , the system exhibits no obvious substructure in simulated $50$ks \Chandra~ images with the system placed at $z=0.1$ (see Poole06a for details on how this is done).

This account of our systems' dynamical evolution is terse and we refer readers to Poole06a for a more detailed and quantitative account of the various processes involved.

\subsection{Some comments regarding the frequency of mergers}\label{numerics-statistics}

In this paper, we seek to understand the role which mergers are likely playing in shaping the statistical properties of clusters as studied through observed scaling relations.  It is important to note however that our simulations certainly are not uniformly represented in these statistics.

In current theories of hierarchical cluster formation, the likelihood of merger events is expected to have a strong dependence on the mass ratio of the interacting systems.  Equal mass mergers are expected to be rare while 10:1 mergers are expected to be common.  Recent studies examining merger statistics in cosmological N-body simulations include \citet{Cohnetal05} and \citet{Wetzeletal06}.  For the simulations presented by \citet{Wetzeletal06}, the fractions of clusters which have experienced 2:1, 3:1 and 10:1 mergers since $z=0.5$ are $23$\%, $49$\% and $92$\% respectively (private communication; from a sample of 37388 clusters with $M>10^{14}M_\odot$ at $z=0$ generated in a $\Lambda CDM$ cosmology with $\Omega_M=0.3$, $h=0.7$, $n=1$ and $\sigma_8=0.9$).  These statistics are generated using a mas jump criteria over conformal time intervals of $\delta \tau=130 h^{-1}$\Mpc.  However, since multiple mergers which occur over this interval will be counted as single events, these numbers are certainly biased high.  These results are supported qualitatively by an analytic extended Press-Schechter analysis we have performed on a set of merger trees generated following the approach of \citet{Tayloretal04}.

Thus, our 3:1 mergers should be representative of the most massive merger events statistically important to the cluster population.  Although there is evidence that equal mass mergers have occurred \citep[\eg\ the ``Cloverleaf cluster'' studied by][]{Finoguenovetal05}, they are rare and we include them in this study primarily as theoretically interesting limiting cases.  The vast majority of clusters should have experienced at least one merger similar to our 10:1 cases.

For these reasons, when considering the effects of mergers on scaling relations, we will primarily be concerned about the effects of 3:1 and 10:1 events.

\section{Luminosity and Temperature}\label{analysis-LT}

\begin{figure*}
\begin{minipage}{175mm}
\begin{center}
\leavevmode \epsfysize=9.8cm \epsfbox{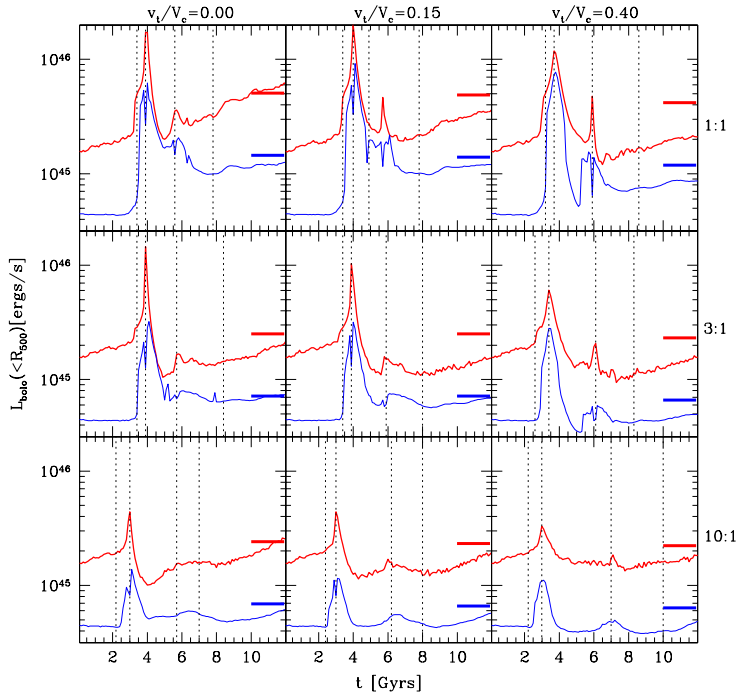}
\caption{Bolometric X-ray luminosities ($L_x$) measured within $R_{500}(t)$.  The thin blue curve traces the evolution of the system with the central $0.1R_{200}$ excised from consideration (\ie~ ``core corrected'' luminosities) while the thick red curve traces the evolution with this region included (\ie~ ``uncorrected'' luminosities).  Thick horizontal dashes indicate the initial bolometric luminosities scaled by the change in total mass within $R_{500}$ from initial to final states using the observed $M_t-L_x$ scaling relation of \citet{ReiprichBohringer02}.  Vertical dotted lines indicate (from left to right) the times $t_o$ (virial crossing), \tclosest~ (first pericenter), \taccrete~ (second pericenter) and \trelax~ (the moment our remnants would appear relaxed to a $50$\ks~ \Chandra~ exposure at $z=0.1$).  Black text around the boundary indicates the mass ratio and $v_t/V_c$ depicted by each panel.}
\label{fig-Lx_t}
\end{center}
\end{minipage}
\end{figure*}

In this section we present the details of how we compute global luminosities and temperatures for our simulations.  We shall find that the temperatures of our merger remnants scale in accord with observed mass scaling relations, but that the luminosity does not.  Excising the cores of our clusters improves this discrepancy.  We will then illustrate the consequences of this for the $L_x-T_x$ scaling relation.

\subsection{Luminosity}\label{analysis-luminosity}

A cluster merger leads to transient increases in both X-ray luminosity ($L_x$) and temperature ($T_x$) which can pose significant problems for cosmological studies \citep{Randalletal02}.  In their study, RS01 found that after an initial transient increase at the time of first pericentric passage, the luminosity of the system drops to a fraction of its initial value with a second smaller peak $\sim 2$\Gyrs~ later when the secondary core returns from apocentric passage.  These sharp transient increases in $L_x$ can translate into a skewing of the high-mass end of mass functions derived from observed X-ray luminosity functions and mass-luminosity scaling relations.  This can translate into a upward bias in $\sigma_8$ of as much as $20$\% and a downward bias in $\Omega_M$ of comparable extent.

In Fig. \ref{fig-Lx_t} we present in red the temporal evolution of the bolometric X-ray luminosities of our simulations integrated within the varying radius$^{7}$\footnotetext[7]{Throughout this paper, we adopt the peak of the projected bolometric X-ray surface brightness as the center of the system.} $R_{500}(t)$.  Although detailed comparisons to RS01 are complicated by the fact that we are not integrating over the entire simulation volume, we see a similar double peaked evolution in luminosity to what is noted in their work.  Furthermore, we find a similar trend in the reduction of the final remnant's luminosity with impact parameter.

This plot illustrates an evolutionary progression with features which we will show are common to all of the global X-ray and SZ observables we study in this paper.  After a short period with very little evolution, the system experiences a dramatic ``first transient'' increase following $t_o$ which reaches maximum amplitude at \tclosest.  This is a product of the increased densities and temperatures resulting from the compressive forces and shocks formed during the secondary's approach to its first pericentric passage through the primary.  Following \tclosest, the system returns to a state similar to it's initial conditions.  Depending on the observable, it can then experience a ``second transient'' due to the recollapse of the system in head-on cases or the second pericentric passage of the surviving portion of the secondary core in off-axis cases.  This second transient is weak for the luminosity of our systems but as we will see, can be more significant for other observables (\eg\ X-ray temperature).  Following the second transient, the system then evolves quiescently as the system relaxes.  This progression is in good qualitative agreement with the results of previous idealized merger studies (\eg\ RS01).  Since all of the X-ray and SZ observables we study have direct dependencies on gas density and temperature, they all evolve with this same qualitative behaviour.  This leads to a similarly generic qualitative evolution (with occasional important differences) in all the scaling relations we will examine.

There is however one notable and generic feature in the luminosity evolution of our simulations not present in RS01: a persistent increase in luminosity after \trelax.  This increase follows the relation $L_x(t)=L_x(t_{relax})\exp{\left( \alpha t \right)}$ where $\alpha=0.08-0.14$\Gyr$^{-1}$.  We also plot, in blue on this figure, the evolution of each system's luminosity with the projected central $0.1R_{200}$ excised (a common measure of the cooling radius of the system; $R_{200}=1785$\kpc and $R_{cool}=180$\kpc~ initially for our primary systems).  These curves (and an examination of the evolving surface brightness profiles of our simulations) reveal that most of this change is arising from increases in the surface brightness of the system within this excised region.  This is a result of increasing central gas densities arising primarily from cooling and (to a lesser extent) the reaccretion of material to the center of the system, dispersed to large radii during the interaction.

\begin{figure*}
\begin{minipage}{175mm}
\begin{center}
\leavevmode \epsfysize=9.8cm \epsfbox{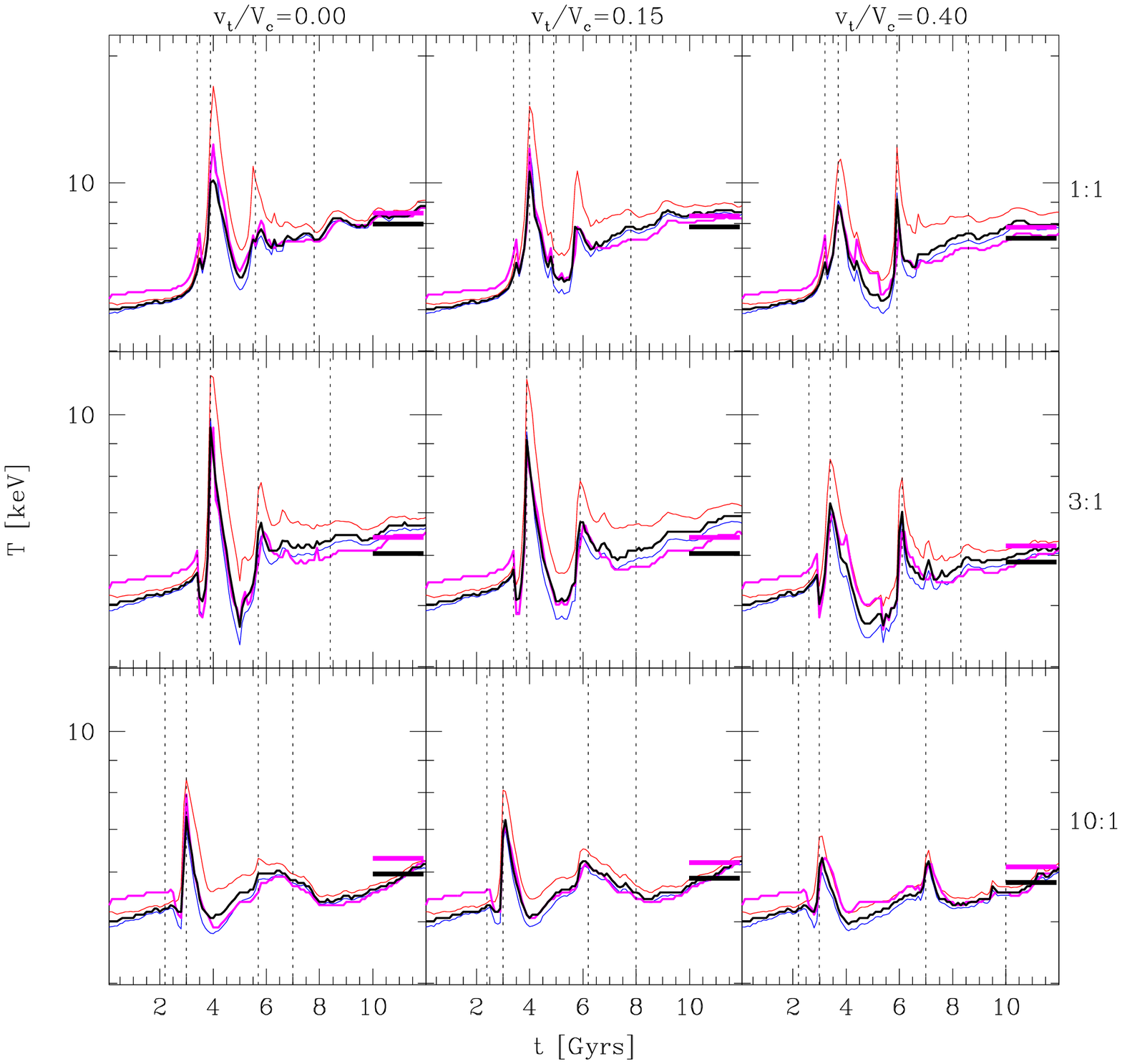}
\caption{Temperatures ($T_x$) measured within $R_{500}(t)$ using several techniques discussed in the text.  Emission-weighted temperatures ($T_{ew}$) are illustrated with red curves, spectrally fit temperatures ($T_{spec}$) with black curves, and ``spectroscopic-like'' \citep[$T_{sl}$, computed according to][]{Mazzottaetal04} temperatures with blue curves.  The thick magenta curves are spectrally fit temperatures with the cool cores ($r<0.1R_{200}$) excised.  Horizontal dashes indicate spectral temperatures scaled from their initial values using the observed $M_t-T_x$ scaling relation from the data of \citet{Horner_thesis} and \citet{ReiprichBohringer02}.  Vertical dotted lines indicate (from left to right) the times $t_o$ (virial crossing), \tclosest~ (first pericenter), \taccrete~ (second pericenter) and \trelax~ (the moment our remnants would appear relaxed to a $50$\ks~ \Chandra~ exposure at $z=0.1$).  Black text around the boundary indicates the mass ratio and $v_t/V_c$ depicted by each panel.  Note the change in vertical scale.}
\label{fig-T_t}
\end{center}
\end{minipage}
\end{figure*}

Throughout this paper we will examine the degree by which our simulated mergers evolve to final states which maintain observed scaling relations.  We do not necessarily expect our simulations to preserve these scaling relations but we use these measures as a means of estimating the degree by which we expect each event to contribute to the observed relation's scatter.  For this reason, we have plotted thick horizontal dashes on Fig. \ref{fig-Lx_t} to indicate the bolometric luminosities we would expect if our remnant systems evolved to states which preserve the observed $M_t-L_x$ scaling relation determined by \citet{ReiprichBohringer02}.  We use the total mass integrated within $R_{500}$ for this purpose.  In head-on cases we see that the total luminosity (in red) of our remnant systems generally manage to recover their expected values by the end of the simulation.  However, the luminosities of our off-axis merger remnants are systematically lower (by $16-50$\%; see table \ref{table-scaling}) than expected from the observed mass scaling relation.  This discrepancy generally increases with impact parameter.  It is consistant with the simulations of \citet{McCarthyetal07} who find that off-axis mergers produce remnant cores with systematically higher entropies and hence, lower densities and luminosities.

Luminosities computed with the central $0.1R_{200}$ excised (in blue) are more successful at preserving the observed mass scaling relation.  The discrepancies of the final luminosities from these scalings are only $3-28$\% (see table \ref{table-scaling}) in these cases.

\subsection{Temperature}\label{analysis-temperature}

Recently several authors have raised concerns that traditional methods of computing ICM temperatures in theoretical studies may be introducing systematic biases during comparisons to observations \citep{MathiesenEvrard01,Gardinietal04,Mazzottaetal04,Vikhlinin06}.  This bias is introduced because the ICM of both theoretical and observed systems is not isothermal and the methods used to compute representative temperatures have been fundamentally different.

Most theoretical measures of temperature are computed through one of a variety of possible weighted averages while observational temperatures are obtained from fits of plasma models to observed spectra.  The most widely used theoretical measure has been the emission-weighted temperature computed from
\begin{center}
\begin{equation}\label{eqn-T_ew}
T_{ew}\equiv \frac{\int \Lambda(T)n_e^2T\,dV}{\int \Lambda(T)n_e^2\,dV}
\end{equation}
\end{center}

\noindent where $\Lambda(T)$ is the bolometric emissivity of the ICM and $n_e$ its electron density.  More accurate but similarly convenient and computationally inexpensive ``spectroscopic-like'' weightings have recently been proposed by \citet{Mazzottaetal04} and \citet{Vikhlinin06}.  From comparisons to hydrodynamic simulations, \citet{Mazzottaetal04}  find that the weighting
\begin{center}
\begin{equation}\label{eqn-T_sl_Mazzotta}
T_{sl}\equiv \frac{\int n_e^2T^{1-\alpha}\,dV}{\int n_e^2T^{-\alpha}\,dV}
\end{equation}
\end{center}

\noindent with $\alpha=0.75$ gives results in much better agreement to temperatures obtained from simulated \Chandra~ observations.  In their analysis, they typically find $T_{sl}$ to be lower than $T_{ew}$ by $20-30$\%.  This relation is calibrated from mixtures of isothermal plasmas of two temperatures, with variations of $T_{sl}$ from spectral fits being less than $10$\%.  

The weighting of \citet{Vikhlinin06} is somewhat more complicated but more accurately takes into account effects on $T_x$ introduced by instrumental variations and metallicity dependent line emission.  They find a very similar power law scaling ($\alpha=0.79$ for \Chandra) as \citet{Mazzottaetal04} for continuum contributions but significant spatial variations may exist between the two measures, particularly in cluster cores which are dominated by low temperature line-emitting gas.  However, we are only interested in globally averaged temperatures in this work, which are dominated by continuum emission from hot gas.  Under these circumstances, very little difference is expected between these measures and so we shall presently focus our attention on the simpler method of \citet{Mazzottaetal04}.

\begin{figure*}
\begin{minipage}{175mm}
\begin{center}
\leavevmode \hspace*{-1cm} \epsfysize=19cm \epsfbox{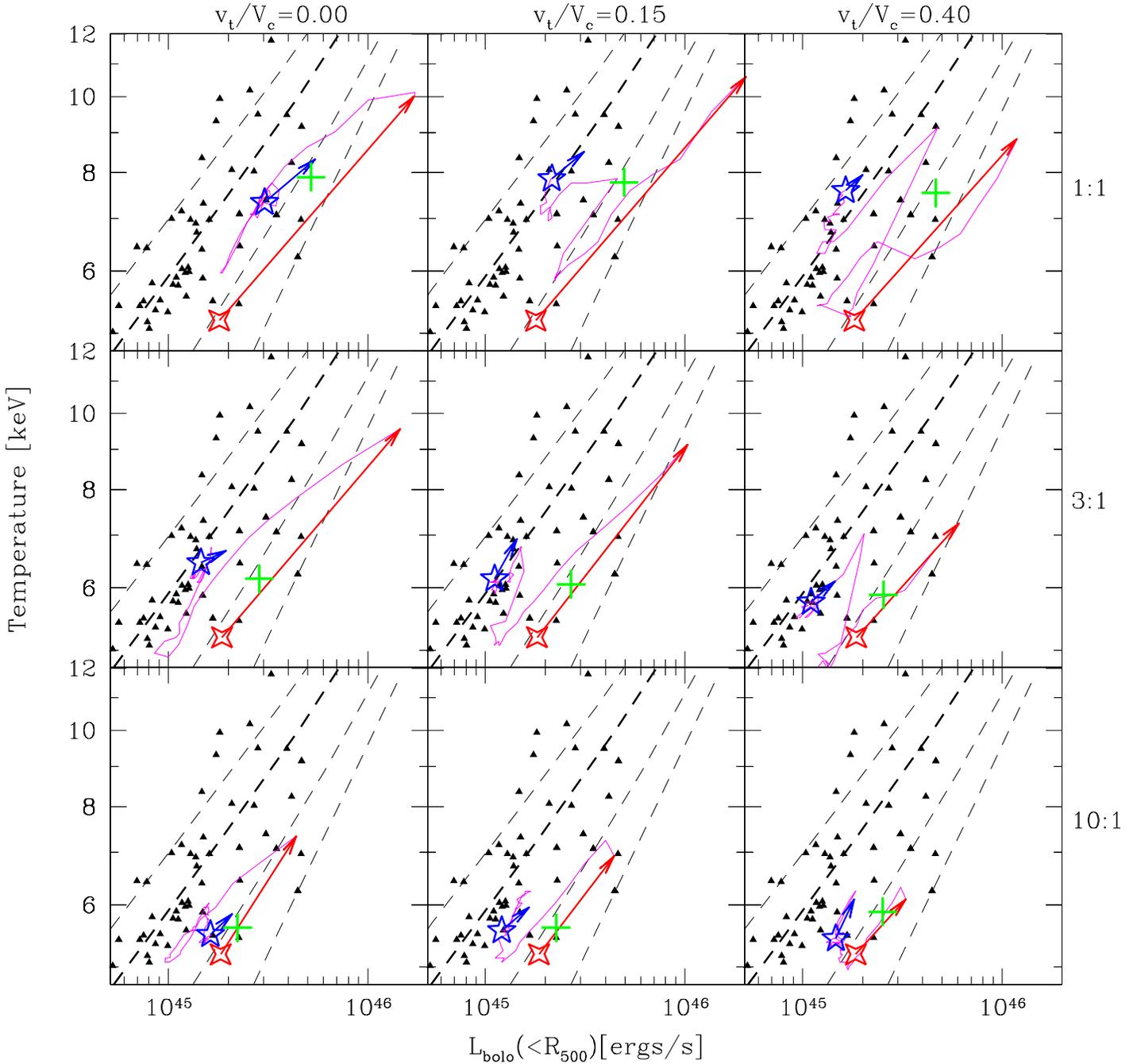}
\caption{X-ray temperatures (spectrally fit) plotted against bolometric luminosities for our simulations (both integrated within $R_{500}$) and compared against observations.  Red four and blue five point stars indicate the states of the system at $t_o$ and \trelax~ respectively.  The red vector indicates the evolution from $t_o$ to \tclosest~ while the blue vector illustrates $3$\Gyr~ of evolution following \trelax.  The magenta line tracks the evolution from \tclosest~ to \trelax~ (generally $4$-$5$\Gyrs) during which the system is visibly disturbed.  Green crosses indicate the remnant state necessary to preserve the observed mass-scaling relations.  Black points are the observed catalogue of \citet{Horner_thesis}.  Dashed lines are our fiducial analytic entropy injection models ($S_o=10$,$100$,$300$, and $500$\keVcmsq, increasing from right to left).  Black text around the boundary indicates the mass ratio and $v_t/V_c$ depicted by each panel.}
\label{fig-L_T}
\end{center}
\end{minipage}
\end{figure*}

Our systems constitute mixtures of gas covering an extremely wide range of densities and temperatures, naturally raising concerns that the specific weightings determined for $T_{sl}$ may breakdown occasionally during the interaction.  To check this we have also computed spectrally fit temperatures ($T_{spec}$).  To produce this measure we compute integrated spectra over the \Chandra~ bandpass ($0.7$-$7$\keV~ with $20$\eV~ resolution) and fit isothermal plasmas to them (we assume a metallicity $Z=0.3Z_\odot$ for all calculations in this paper).  This is done via $\chi^2$ minimization weighted by the photon flux of each spectral bin.  The instrumental response of \Chandra~ is included in this analysis as well as Galactic absorption at a level of $N_H=2\times 10^{20}cm^{-2}$.

In the analysis of RS01, the emission-weighted temperature of the system remains relatively constant, interrupted only by the pair of transients discussed above, during first and second pericentric passages when the cluster cores are shock heated and compressed.  As their systems relax, they recover to slightly higher emission-weighted temperatures in equal mass events and to approximately the initial temperature in 3:1 cases.

In Fig. \ref{fig-T_t} we present the temporal evolution of the three temperature measures discussed above, integrated within $R_{500}(t)$ for our simulations.  In all cases we witness the familiar double peaked behaviour noted by RS01.  We additionally observe the trend noted by RS01 that the amplitude of the initial peak decreases with impact parameter while the relative strength of the two peaks becomes more similar.  

It has been noted by \citet{Mazzottaetal04} that the mismatch between $T_{ew}$ and $T_{sl}$ increases with the thermal complexity of the system.  This is reflected in our simulations by an increased discrepancy between $T_{ew}$ and our other temperature measures during the transient temperature increases at first and second pericentric passage.  This discrepancy, which is $15-40$\% larger than the difference between $T_{ew}$ and $T_{sl}$ at all other times, suggests that the transient increases determined by RS01 (who present emission-weighted temperatures) are significantly overestimated.  Consequently it suggests that the influence of transient temperature increases on measures of $\sigma_8$ determined by \citet{Randalletal02} are overestimated.

As expected, $T_{sl}$ and $T_{spec}$ are systematically lower than $T_{ew}$ at all times.  Initially, this difference is small ($\sim 3$\%) but at late times it can be as much as $\sim 10$\%.  Given that $T_{sl}$ produces results in very good agreement with $T_{spec}$ throughout our merger interactions (less than $5$\% difference in all cases and at all times except $500$\Myrs~ during first pericenter in the 1:1 near-axis cases where we see a $20$\% difference), we conclude that the methods of spectrally weighted temperature averages developed by \citet{Mazzottaetal04} and \citet{Vikhlinin06} are indeed robust during the complicated evolution of a cluster merger.  We shall use our spectrally fit temperatures for the remainder of our analysis however, unless otherwise stated.

Given that there is such a persistent offset between $T_{ew}$ and $T_{spec}$, the normalization of theoretical temperature scaling relations generated with this approach are likely too high.  This has important implications for cosmological studies utilizing emission-weighted temperature functions as well.

In Fig. \ref{fig-T_t} we place thick horizontal dashes to indicate the spectral temperature the final remnant would have if it were scaled from the initial temperature by the $M_t-T_x$ relation derived from the temperatures of \citet{Horner_thesis} and masses of \citet{ReiprichBohringer02}.  We see that immediately following \trelax, the system is significantly cooler than this scaled value.  Late increases evolve according to $T_x(t)=T_x(t_{relax})\exp{\left( \alpha t \right)}$ (with $\alpha=0.01-0.03$\Gyr$^{-1}$), taking the system to roughly the observed mass-scaled temperature by the ends of our simulations in all cases.  Of the three temperature measures, $T_{ew}$ produces results which most often and most significantly fail to yield the mass-scaled result, suggesting that previous theoretical studies of the $L_x-T_x$ and $M_t-T_x$ relations which have employed $T_{ew}$ as a temperature measure have exaggerated the scatter in this relation introduced by mergers.  In Table \ref{table-scaling}, we list the discrepancies of our remnants' temperatures from the observed mass-scaling relation.

\subsection{$L_x-T_x$ relation}\label{analysis-L_T}

In Fig. \ref{fig-L_T} we present the evolution of our systems in the $L_x-T_x$ plane.  We illustrate the evolution from the initial state (red 4 point star) to \tclosest~ (first pericentric passage) with a red arrow and the evolution for $3$ \Gyrs~ following \trelax~ (blue five point star) with a blue arrow.  In each case, the final $L_x$ and $T_x$, scaled from their initial values by the observed mass scaling relations of \citet{ReiprichBohringer02} and \citet{Horner_thesis}, are illustrated by a green cross.  Magenta curves track the evolution of the system from \tclosest~ to \trelax, while the system appears significantly disturbed (an interval of $4-5$\Gyrs, except for the 10:1 $v_t/V_c=0.4$ case which takes $7$\Gyrs~ to appear relaxed).  This format will be used for all the scaling relations we subsequently present.

It has been known for some time that variations in the structure of the central regions of clusters contribute significantly to the scatter in both luminosity and temperature mass scaling relations.  For this reason, and out of a desire to obtain the tightest relations possible (usually for the purposes of obtaining mass functions for cosmological studies), authors have typically presented ``core-corrected'' observations for studying the $L_x-T_x$ relation.  It has been shown by M04 however that there is very interesting structure in the $L_x-T_x$ plane which correlates with the morphology of the cores when the central regions are retained for analysis.  To explore the consequences (and possible contributions of) mergers to this structure, we use the ``uncorrected'' catalogue presented by \citet{Horner_thesis} for comparison of our results to observations (presented in black on Fig. \ref{fig-L_T}).

On Fig. \ref{fig-L_T} (and most subsequent scaling relation plots) we have also placed dashed lines depicting theoretical mass scaling relations for a set of fiducial analytic  entropy injection models for systems with various minimum entropies.  These models are computed following the procedure of \citet{MBBPH} with the following exceptions: we use the mass-concentration relation from the Millennium Simulation, the baseline entropy profile of \citet{Voitetal05}, and assume a bias $M_g(<R_{200})/M_t(<R_{200})=0.9 \Omega_b/\Omega_m$.  Four lines are plotted for core entropies of $10$, $100$, $300$ and $500$ \keVcmsq~ (increasing from right to left).  In \citet{BBLP} it was found that the $300$\keVcmsq~ model best fits the median relation of the data.  This model is thus indicated with a thicker line type and we shall consider it to represent the median relation of the data in all discussions which follow.

\begin{figure*}
\begin{minipage}{175mm}
\begin{center}
\leavevmode \epsfysize=9.8cm \epsfbox{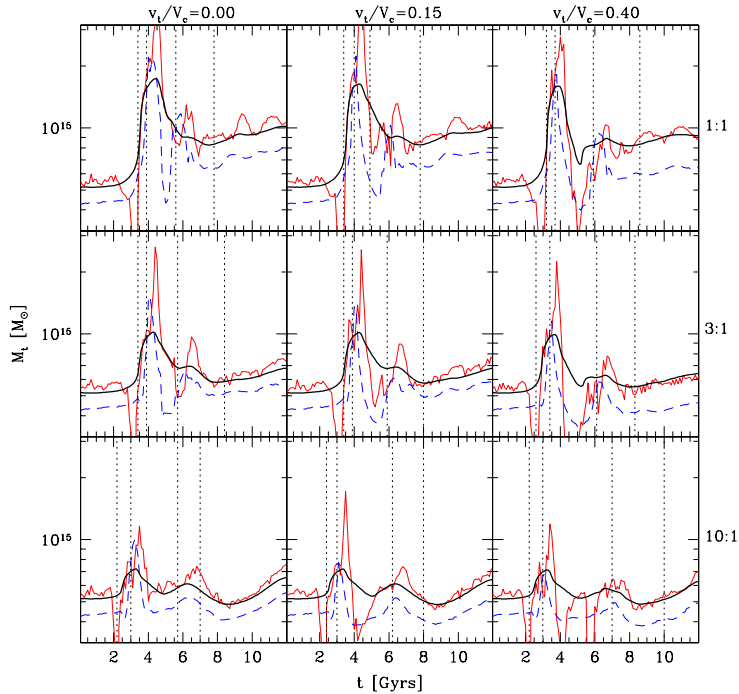}
\caption{Total masses within $R_{500}(t)$ computed from our simulations.  The actual total mass is plotted as a thick black curve, the hydrostatic equilibrium mass (computed from Eqn.\ref{eqn-hydro_eqm}) is plotted as a red curve and the isothermal $\beta$-model mass (computed from Eqn.\ref{eqn-M_isobeta}) is plotted as a dashed blue curve.  Vertical dotted lines indicate (from left to right) $t_o$ (the time of virial crossing), \tclosest~ (first pericenter), \taccrete~ (second pericenter) and \trelax~ (the moment our remnants would appear relaxed to a $50$\ks~ \Chandra~ exposure at $z=0.1$).  Black text around the boundary indicates the mass ratio and $v_t/V_c$ depicted by each panel.}\label{fig-M_t}
\end{center}
\end{minipage}
\end{figure*}

In Section \ref{analysis-luminosity} we discussed the common temporal evolution of the global X-ray and SZ observables we study in this paper.  We also noted that this leads to a common qualitative evolution in their scaling relations as well.  We see the first illustration of this progression in Fig. \ref{fig-L_T}.  Initially, due to the shocks and compressive forces generated as the interacting systems reach first pericenter, the system's luminosity and temperature both increase sharply (illustrated by red vectors).  In all cases, these vectors run roughly parallel to the mass scaling relations of our fiducial models in this plane, supporting similar findings in previous merger studies \citep[\eg~ ][]{RTK04}.  This will be the general case (with some important differences) in the other scaling relations we shall study as well.

Following the first transient, the system returns to nearly it's initial state at $t_o$ and subsequently evolves erratically until \trelax~ (when the system appears relaxed in our simulated \Chandra~ observations and is virialized within $R_{500}$).  The magenta curve traces the system's evolution during this period, which lasts $4-5$\Gyrs.  During this time it experiences a second transient (illustrated by a sharp jump along the fiducial models in the magenta curve) when the system recollapses in the head-on cases or the surviving portion of the secondary's core returns for a second pericentric passage in the off-axis cases.  The magnitude of this second transient increases with impact parameter due to the reduced disturbance of the secondary core.  It is during this period when most of the displacement across the scatter of the observations occurs.  Although difficult to illustrate with this plot, much this displacement occurs during the accretion of streams following the secondary's second pericentric passage in several cases.  We will examine in more detail the complicated processes driving this evolution in \citet{P07a} where we will examine the effects of mergers on the cores of our systems.

Once second pericentric passage has passed and obvious substructure dissolved, the blue vector (which illustrates $3$\Gyrs\ of evolution following \trelax) shows that the system once again evolves parallel to the mass scaling relations of our fiducial models.  This occurs as cooling leads to a denser, more luminous core and material dispersed by the interaction reaccretes to the system centre, deepening its gravitational potential and increasing its temperature.

The initial conditions we have chosen place our primary systems on the high luminosity side of the scatter in the $L_x-T_x$ plane, where systems with compact cool cores are typically found (M04).  In Section \ref{analysis-temperature} we found that mergers tend to evolve to states which preserve the observed mass-temperature scaling relation but in Section \ref{analysis-luminosity} found that they create remnants which are under luminous with respect to the observed mass-luminosity scaling relation.  In Fig. \ref{fig-L_T} we see the consequences of these trends: off-axis mergers tend to push typical compact cold core systems towards the low-luminosity side of the observed $L_x-T_x$ relation.

In our off-axis 1:1 cases (which we must emphasize, are rarely observed and expected to be exceptionally uncommon) our mergers are able to drive systems to states which cover the majority of the scatter in this plane.  However, in the more common situations represented by our 3:1 mergers, states represented by our fiducial entropy injection models with entropies higher than $S=300$\keVcmsq~ are not reached, not even transiently.  Our 10:1 mergers fail to reach the median relation of the data under any circumstance.

When this plot is generated utilizing emission-weighted temperatures, the full dispersion of the observations is covered by both our 1:1 and 3:1 mergers.  It is now well established that this temperature measure is not viable however, so this does not represent a means by which to account for the observed scatter.  Instead, it further emphasizes the importance of computing proper spectral temperatures when comparing theoretical models of X-ray clusters to observations.  Furthermore, not only should the normalization of past temperature scaling relations generated with the emission-weighted measure be viewed with some scepticism (as argued in Section \ref{analysis-temperature}), the scatter should be as well.

\section{Mass}\label{analysis-mass}

To date, all catalogues of cluster masses involving significant sample sizes have been compiled from X-ray observations.  The methods used are thus not direct (such as with lensing measures) but are inferred from temperature and surface brightness profiles using methods which may be highly susceptible to systematic biases, particularly during a merger.  These approaches generally assume that the system can be described as being in a state of hydrostatic equilibrium; a condition given by
\begin{center}
\begin{equation}
\rho_g^{-1}\frac{dP}{dr}=-\frac{GM_t(<r)}{r^2}
\label{eqn-hydro_eqm_def}
\end{equation}
\end{center}
\noindent This yields a mass profile of the general form
\begin{center}
\begin{equation}
M(r)=-\frac{kT(r)}{G\mu m_p} r \left[ \frac{d \log \rho_g(r)}{d \log r} + \frac{d \log T(r)}{d \log r} \right]
\label{eqn-hydro_eqm}
\end{equation}
\end{center}
\noindent where $G$ and $k$ are Newton's and Boltzmann's constants respectively, $T(r)$ is the temperature profile of the system, $\mu$ the mean molecular weight, $m_p$ the mass of the proton and $\rho_g(r)$ the gas density profile.
In Poole06a we examined the evolution of a hydrostatic disequilibrium parameter given by
\begin{center}
\begin{equation}\label{eqn-H_parameter}
H=1+\frac{\rho_g^{-1}\frac{dP}{dr}}{r^{-2} G M_t(<r)}
\end{equation}
\end{center}
\noindent and illustrated that clusters which appear relaxed can violate the condition of hydrostatic equilibrium by $H=10-15$\% and occasionally even more.  How does this affect masses obtained from Eqn. \ref{eqn-hydro_eqm}?

In Fig. \ref{fig-M_t} we present (in red) the evolution of the mass of our systems within $R_{500}(t)$ computed directly from our simulations using this relation.  For comparison, the actual total mass is presented as well (thick black).  We can see from this plot that despite the limited validity of the hydrostatic equilibrium condition, masses computed from Eqn. \ref{eqn-hydro_eqm} after \trelax~ (when the system appears as a single undisturbed remnant) generally reproduce the system mass to within $\sim 5-10$\%, with a slight systematic bias towards higher values.

To implement Eqn. \ref{eqn-hydro_eqm} to measure cluster masses without additional assumptions, accurate measurements of $\rho_g(r)$ and $T(r)$ must be made.  The observational challenges of doing so have been considerable but with the increasing availability of high quality \Chandra~ and \XMM~ observations, significant progress has recently been made.  \citet{Vikhlininetal06} for instance have presented a method of gas density and temperature deprojection which they have demonstrated to be accurate to a few percent for apparently relaxed systems \citep{Nagaietal06}.  

Presently however, statistically representative \Chandra~ and \XMM~ catalogues of sufficient size to study mass scaling relations are not available.  For this reason, we shall use for comparison the catalogue of \citet{ReiprichBohringer02} who report masses within $R_{500}$ measured from ROSAT and ASCA observations.  These authors measure masses assuming that their observed systems are in hydrostatic equilibrium but make the further assumptions that the gas is isothermal and that it's density profile follows a parameterized form given by the $\beta$-model.  For an X-ray surface brightness profile given by
\begin{center}
\begin{equation}
S_x(r_p)=\frac{S_o}{\left( 1 +(r_p/r_c)^2 \right)^{(6\beta - 1)/2}}
\label{eqn-beta_model_Sx}
\end{equation}
\end{center}

\noindent the gas density profile of an isothermal gas is given by
\begin{center}
\begin{equation}
\rho_g(r)=\frac{\rho_o}{\left( 1 +(r/r_c)^2 \right)^{3\beta/2}}
\label{eqn-beta_model_rho}
\end{equation}
\end{center}

\noindent Under these assumptions, fitting Eqn. \ref{eqn-beta_model_Sx} to the surface brightness profile yields the logarithmic derivative of the density profile, producing the following relation for the total mass within $R_{500}$
\begin{center}
\begin{equation}
M_{\beta}(R_{500})=\frac{3\beta k T_x R_{500}}{G\mu m_p} \frac{(R_{500}/r_c)}{1+(R_{500}/r_c)^2}
\label{eqn-M_isobeta}
\end{equation}
\end{center}

\noindent Temperature and $\beta$-model fits in the results presented by \citet{ReiprichBohringer02} are fit to the flux which extends to the observed radius of the system (with the central $0.1R_{200}$ excised for the temperature fits).  The observed radius is generally comparable to $R_{500}$ which is the radius we use to generate the isothermal $\beta$-model masses (computed from Eqn. \ref{eqn-M_isobeta}) we present in Fig. \ref{fig-M_t} (in blue).

From this figure we can see that the isothermal $\beta$-model systematically underestimates the mass of our systems by $25-40$\%.  This result is in good agreement with the findings of several other authors \citep[\eg~][]{Hallmanetal06,Rasiaetal06,Kayetal04}.  The general trends in the scaling relations should not be affected much by this bias, but the normalization will be.  This could have significant implications for cosmological studies conducted with isothermal $\beta$-model masses.

\subsection{$M_t-T_x$ relation}\label{analysis-M_T}

\begin{figure*}
\begin{minipage}{175mm}
\begin{center}
\leavevmode \hspace*{-1cm} \epsfysize=19cm \epsfbox{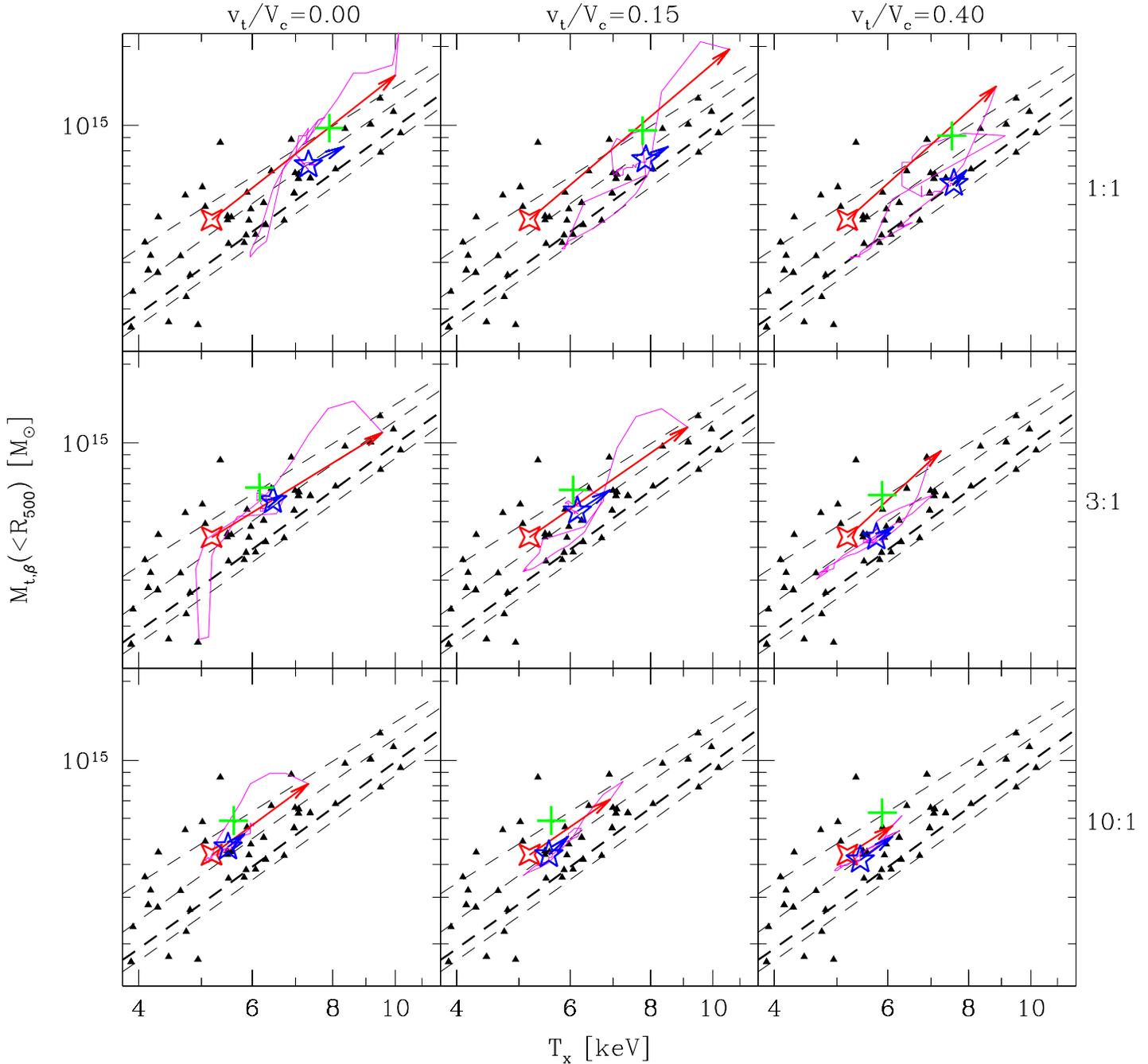}
\caption{Isothermal $\beta$-model masses plotted against spectrally fit temperatures for our simulations (both integrated within $R_{500}$) and compared against observations.  Red four and blue five point stars indicate the states of the system at $t_o$ and \trelax~ respectively.  The red vector indicates the evolution from $t_o$ to \tclosest~ while the blue vector illustrates $3$\Gyr~ of evolution following \trelax.  The magenta line tracks the evolution from \tclosest~ to \trelax~ (generally $4$-$5$\Gyrs) during which the system is visibly disturbed.  Green crosses indicate the remnant state necessary to preserve the observed mass-scaling relations.  Black points are the observed catalogues of \citet{Horner_thesis} (temperatures) and \citet{ReiprichBohringer02} (masses).  Dashed lines are our fiducial analytic entropy injection models ($S_o=10$,$100$,$300$, and $500$\keVcmsq, increasing from left to right) with the masses reduced by $30$\% to account for the bias in the isothermal $\beta$-model approach.  Black text around the boundary indicates the mass ratio and $v_t/V_c$ depicted by each panel.}\label{fig-M_T}
\end{center}
\end{minipage}
\end{figure*}
\begin{figure*}
\begin{minipage}{175mm}
\begin{center}
\leavevmode \hspace*{-1cm} \epsfysize=19cm \epsfbox{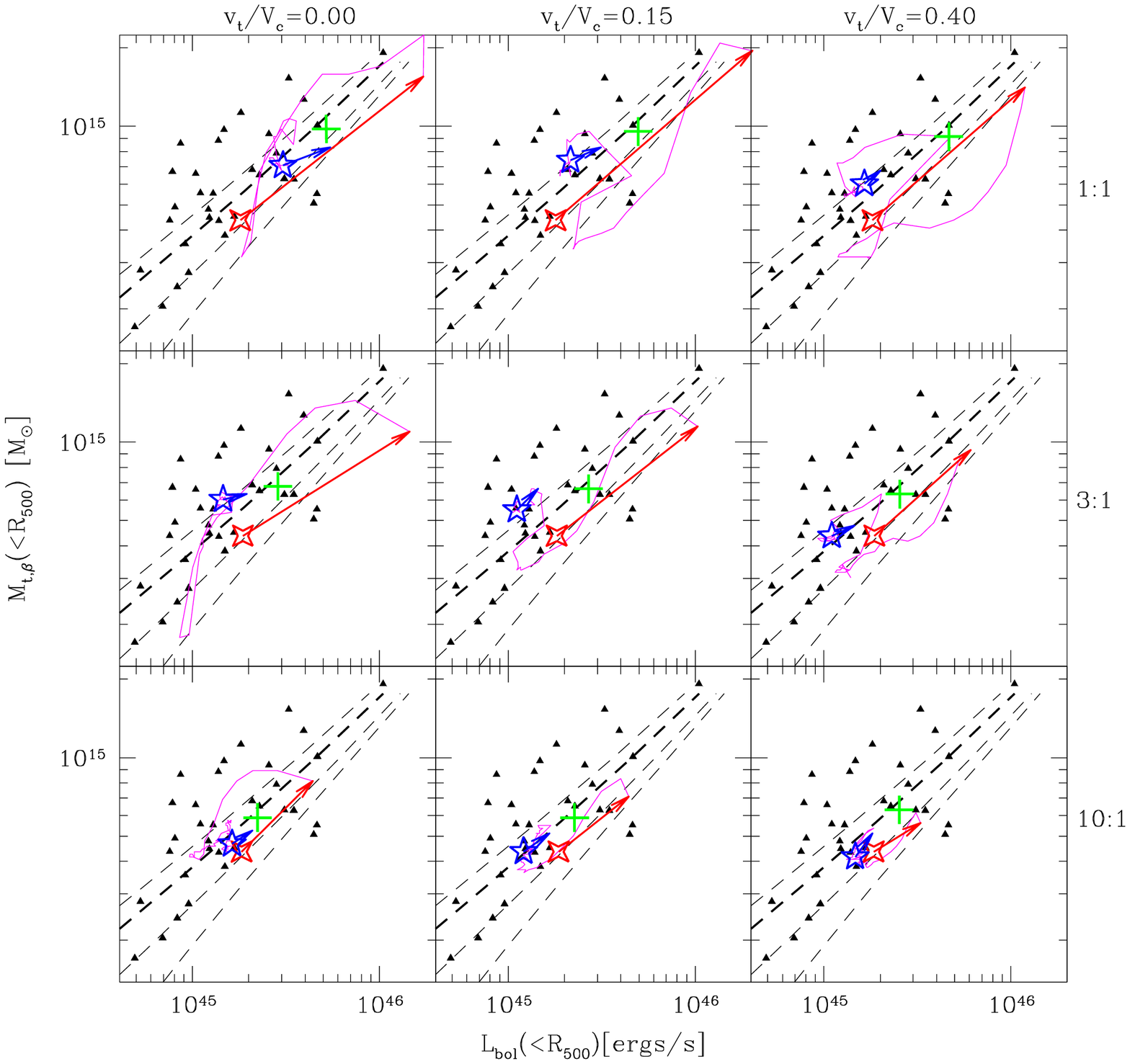}
\caption{Isothermal $\beta$-model masses plotted against bolometric luminosities for our simulations (both integrated within $R_{500}$) and compared against observations.  Red four and blue five point stars indicate the states of the system at $t_o$ and \trelax~ respectively.  The red vector indicates the evolution from $t_o$ to \tclosest~ while the blue vector illustrates $3$\Gyr~ of evolution following \trelax.  The magenta line tracks the evolution from \tclosest~ to \trelax~ (generally $4$-$5$\Gyrs) during which the system is visibly disturbed.  Green crosses indicate the remnant state necessary to preserve the observed mass-scaling relations.  Black points are the observed catalogue of \citet{ReiprichBohringer02}.  Dashed lines are our fiducial analytic entropy injection models ($S_o=10$,$100$,$300$, and $500$\keVcmsq, increasing from right to left) with the masses reduced by $30$\% to account for the bias in the isothermal $\beta$-model approach.  Black text around the boundary indicates the mass ratio and $v_t/V_c$ depicted by each panel.}\label{fig-M_L}
\end{center}
\end{minipage}
\end{figure*}

In Fig. \ref{fig-M_T} we present the evolution of our isothermal $\beta$-model masses against temperature and compare them to the observed temperature catalogue of \citet{Horner_thesis} and mass catalogue of \citet{ReiprichBohringer02}.  We follow the same format here as for Fig. \ref{fig-L_T}.  Red four and blue five point stars indicate the system's state initially and when it appears relaxed at \trelax.  The red vector illustrates the transient evolution from $t_o$ to \tclosest~ (first pericentric passage) and the blue vector shows $3$\Gyrs~ of evolution following \trelax.  The magenta line traces the evolution of the system between \tclosest~ and \trelax, when the system looks visibly disturbed.  Green crosses indicate the $\beta$-model masses of our remnants scaled from their initial values by the actual change of mass within $R_{500}$, and their temperatures similarly scaled by the observed $M_t-T_x$ scaling relation.  Dashed lines indicate our fiducial entropy injection models (with the entropy minimum increasing with temperature at a constant mass).  For this plot, we have reduced the masses of these models by $30$\% to reflect the downward bias in isothermal $\beta$-model masses found in Section \ref{analysis-mass}.  When we do so, we find a good agreement between our fiducial models and the observed catalogue.  Since the observed catalogue and our simulation masses are compiled using the same $\beta$-model techniques, they should be directly comparable and do not require any such adjustment.

First, we see that there is a considerably smaller degree of scatter (in both the observations and our fiducial models) in this plane compared to what is seen in the $L_x-T_x$ plane.  This is not a surprise given the strong correlation we expect between these quantities: it is the depth of the potential (which is set by the mass) which determines the virial temperature (and hence, X-ray temperature) of the system.  For the observations, another (potentially more significant) reason for reduced scatter is the $M_{\beta}\propto T_x$ dependence present in the isothermal $\beta$-model.  Any transient increase in $T_x$ is matched by a proportional increase in $M_\beta$, reducing the observed scatter in this plane.

Next we see that the qualitative evolution of our mergers in this plane follows the generic pattern we have discussed in Sections \ref{analysis-luminosity} and \ref{analysis-L_T}: two transient increases in both quantities at \tclosest~ (first pericenter; illustrated by the red vector) and \taccrete~ (second pericenter; illustrated by the magenta curve) which take the system roughly along our fiducial mass scaling relations, followed by a period of relaxation (illustrated by the blue vector) during which the system slowly evolves upward along the scaling relation.  While the system is disturbed (magenta curve), it drifts slightly across the scatter in the observations.  In both planes, the initial transient tends to pull the system towards the low temperature side of the scatter (\ie\ away from the side preferentially occupied by NCC systems).

There are interesting differences between the evolution in the $L_x-T_x$ and $M_t-T_x$ planes however.  Most notably, our mergers occasionally (in the head-on and $v_t/V_c=0.15$ 1:1 and 3:1 cases) spend very brief ($t<0.5$\Gyrs) periods outside of the bounds of our fiducial models, at the very limits of the scatter in the observations.  These periods correspond to those in Fig. \ref{fig-M_t} during which we witness strong failures of the isothermal $\beta$-model at capturing the total mass of the system.  We also see a couple of brief ($t<0.5$\Gyrs) excursions to the high temperature side of the observations after \taccrete~ and before \trelax~ in the 1:1 off-axis and 3:1 $v_t/V_c=0.15$ cases.  These are not seen in the $L_x-T_x$ plane.  This highlights the vulnerability of isothermal $\beta$-model mass estimates to the effects of mergers and the spurious contributions to scatter in the scaling relations likely being introduced by its use.

Such rare differences aside, our simulations suggest that mergers have a similar effect on the scatter of the $M_t-T_x$ plane as on the $L_x-T_x$ plane, for relatively relaxed systems.  Our (exceptionally rare) 1:1 off-axis mergers produce relaxed states which spread across the majority of the scatter in the observations while more common 3:1 and 10:1 mergers do not create states which reach the high central entropy side of the observations preferentially inhabited by NCC systems.

Recently, \citet{Baloghetal06} have studied the $M_t-L_x$ and $M_t-T_x$ scaling relations, paying attention to the scatter introduced by variations in dark matter halo concentrations and formation epochs.  They find that, although reasonable variations in these properties can account for the observed scatter in the $M_t-T_x$ plane, they can not do so for the observations in the $M_t-L_x$ plane.

Variations in dark matter halo concentrations are a product of system-to-system variations in formation histories, which regularly consist of several merger events.  The fact that our simulations generally remain well within the scatter of the $M_t-T_x$ plane lends support to the claims of \citet{Baloghetal06} that mergers do not produce significant scatter beyond their effects in driving variations in formation time and dark matter concentration.

\subsection{$M_t-L_x$ relation}\label{analysis-M_L}

In Fig. \ref{fig-M_L} we present the evolution of our isothermal $\beta$-model masses against luminosity, following the same format as Fig. \ref{fig-M_T}.  Once again we see the same qualitative evolution witnessed in the $L_x-T_x$ plane: two transients at \tclosest~ (first pericentric passage; red vector) and \taccrete~ (second pericentric passage; magenta curve) followed by a period of relaxation (blue vector) during which the system evolves up the mass scaling relations.  Furthermore, we see the same brief failures of the $\beta$-model mass estimates seen in the $M_t-T_x$ plane, which drive the system briefly ($<0.5$\Gyrs) outside the region encompassed by our fiducial analytic models.

The scatter in this plane is much larger and it is clear that our simulations fail to produce states of sufficiently low luminosity to account for the observations in this plane.  There is a hint that our systems evolve towards the low-luminosity side of the observed scaling relation, but only briefly ($<0.5$\Gyrs) and during the extreme disruptions generated during the head-on cases.  In the more common off-axis cases, the systems tend to evolve away from this side of the scatter when significantly disturbed.

\section{Sunyaev-Zel'dovich effect}\label{analysis-SZ}

Quite some time has passed since the original realization that the spectral signature of inverse Compton scattering of the cosmic microwave background (CMB) by the hot ICM of galaxy clusters could be detected as a temperature increment in the CMB and used as a probe of the ICM's structure \citep{SZ70,SZ72}.  Known as the thermal Sunyaev-Zel'dovich (SZ) effect,  the observational challenges involved in its detection have been significant, making abundant and reliable data slow to emerge.

Several sensitive large-area SZ surveys will soon be available however and due to the weak redshift dependence on the mass threshold of a cluster's SZ detectability \citep[$\sim10^{14}$\msun, set by confusion limits; see][for more details]{Holderetal07}, catalogues of tens of thousands of clusters extending to high redshifts will soon be available as an incredibly powerful tool for studying cluster evolution and cosmology.  The potential for mergers to introduce biases in these catalogues has a similar, but likely less severe potential to introduce systematic uncertainties into the analysis of these datasets \citep{MajumdarandMohr03}.  Furthermore, because the SZ effect is dependent on a line of sight integral involving a single factor of density (rather than the $\rho^2$ dependence of X-ray emission studies), SZ observations of mergers will be useful as a complementary tool to X-ray emission observations for determining the structure of the ICM.

\begin{figure*}
\begin{minipage}{175mm}
\begin{center}
\leavevmode \epsfysize=8.9cm \epsfbox{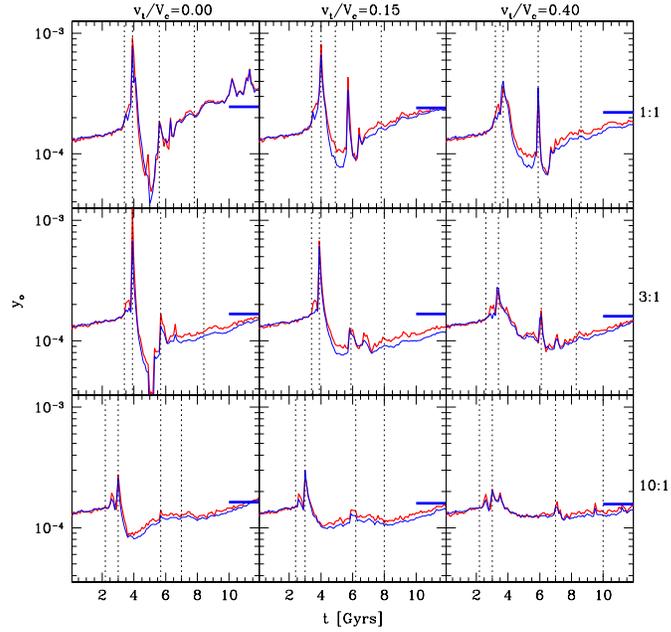}
\caption{The evolution of the central Compton parameter ($y_o$) as a function of time for our simulations.  Results computed assuming an isothermal $\beta$-models are shown with thin blue curves while direct integrations of our simulations (with densities and temperatures within $30$\kpc~ held constant) are shown with thick red curves.  Horizontal dashes represent the final values expected for the $\beta$-model results from the mass scalings derived from our fiducial models (see Section \ref{analysis-luminosity}).  Vertical dotted lines indicate (from left to right) $t_o$ (the time of virial crossing), \tclosest~ (first pericenter), \taccrete~ (second pericenter) and \trelax~ (the moment our remnants would appear relaxed to a $50$\ks~ \Chandra~ exposure at $z=0.1$).  Black text around the boundary indicates the mass ratio and $v_t/V_c$ depicted by each panel.}
\label{fig-yo_t}
\end{center}
\end{minipage}
\end{figure*}
\begin{figure*}
\begin{minipage}{175mm}
\begin{center}
\leavevmode \epsfysize=8.9cm \epsfbox{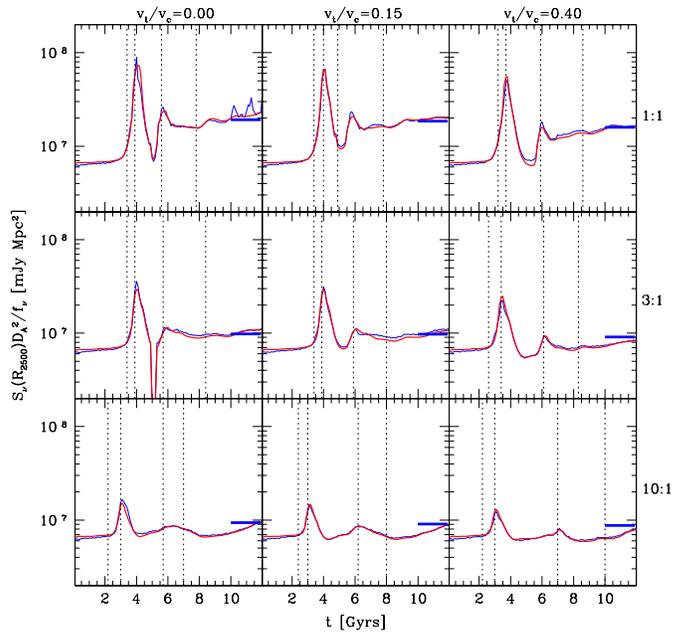}
\caption{The evolution of integrated SZ fluxes ($S_\nu D_A^2/f_\nu$) computed within $R_{2500}$ for our simulations as a function of time.  Results computed assuming an isothermal $\beta$-model are shown with thin blue curves while direct integrations from our simulations are shown with thick red curves.  Horizontal dashes represent the final values expected from the mass scalings derived from our fiducial models (see Section \ref{analysis-luminosity}).  Vertical dotted lines indicate (from left to right) $t_o$ (the time of virial crossing), \tclosest~ (first pericenter), \taccrete~ (second pericenter) and \trelax~ (the moment our remnants would appear relaxed to a $50$\ks~ \Chandra~ exposure at $z=0.1$).  Black text around the boundary indicates the mass ratio and $v_t/V_c$ depicted by each panel.}
\label{fig-Snu_t}
\end{center}
\end{minipage}
\end{figure*}

\begin{figure*}
\begin{minipage}{175mm}
\begin{center}
\leavevmode \hspace*{-1cm} \epsfysize=19cm \epsfbox{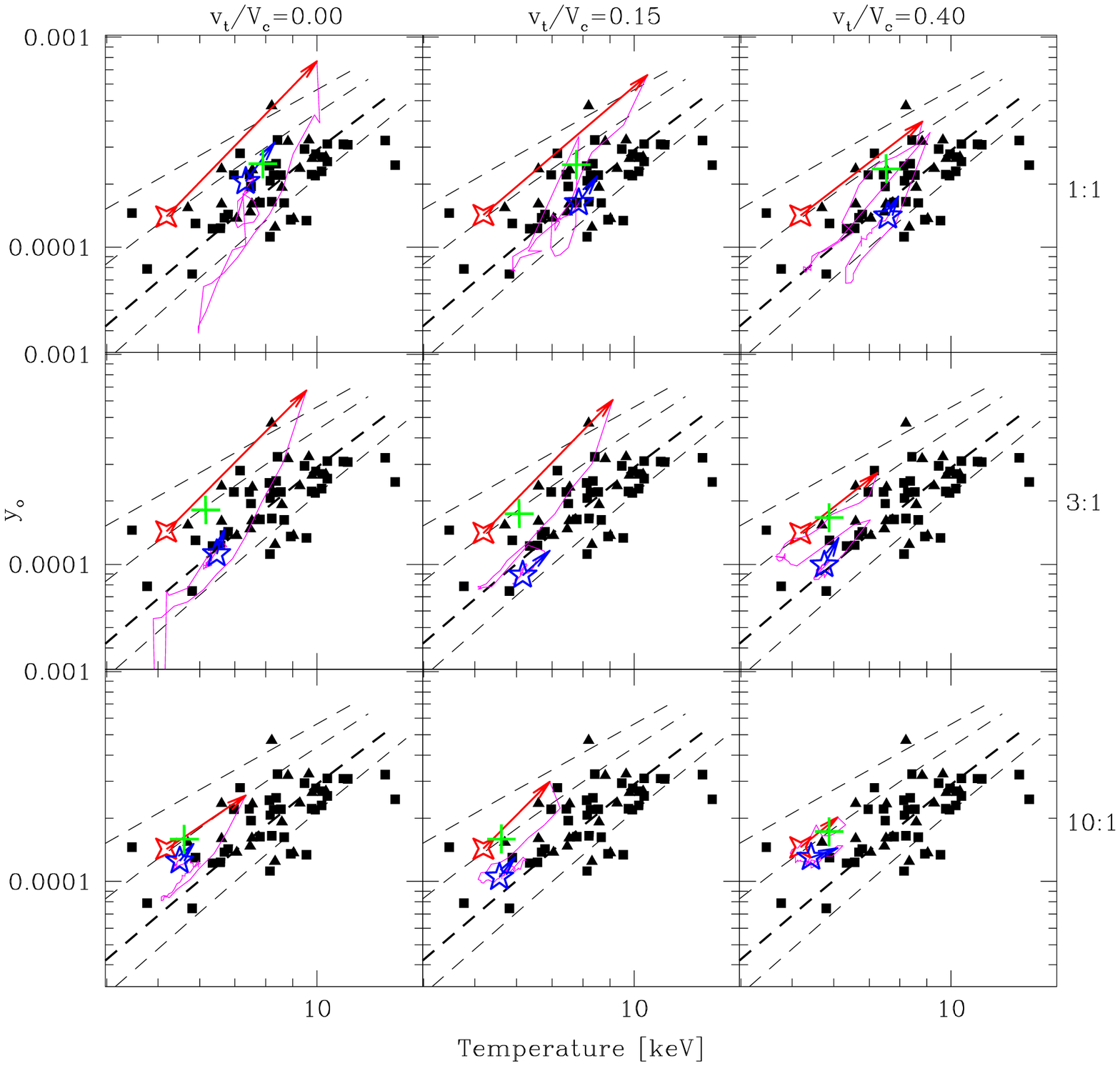}
\caption{Evolution of the SZ Compton parameter ($y_o$) plotted against temperature for our simulations (both integrated within $R_{500}$) and compared against observations.  Red four and blue five point stars indicate the states of the system at $t_o$ and \trelax~ respectively.  The red vector indicates the evolution from $t_o$ to \tclosest~ while the blue vector illustrates $3$\Gyr~ of evolution following \trelax.  The magenta line tracks the evolution from \tclosest~ to \trelax~ (generally $4$-$5$\Gyrs) during which the system is visibly disturbed.  Green crosses indicate the remnant state necessary to preserve the observed mass-scaling relations.  Black points illustrate the observed catalogues of \citet[][triangles]{Reeseetal02} and \citet[][squares]{LaRoqueetal06} with temperatures taken from \citet{Horner_thesis}.  Dashed lines are our fiducial analytic entropy injection models ($S_o=10$,$100$,$300$, and $500$\keVcmsq, increasing from left to right).  Black text around the boundary indicates the mass ratio and $v_t/V_c$ depicted by each panel.}
\label{fig-yo_Tx}
\end{center}
\end{minipage}
\end{figure*}
\begin{figure*}
\begin{minipage}{175mm}
\begin{center}
\leavevmode \hspace*{-1cm} \epsfysize=19cm \epsfbox{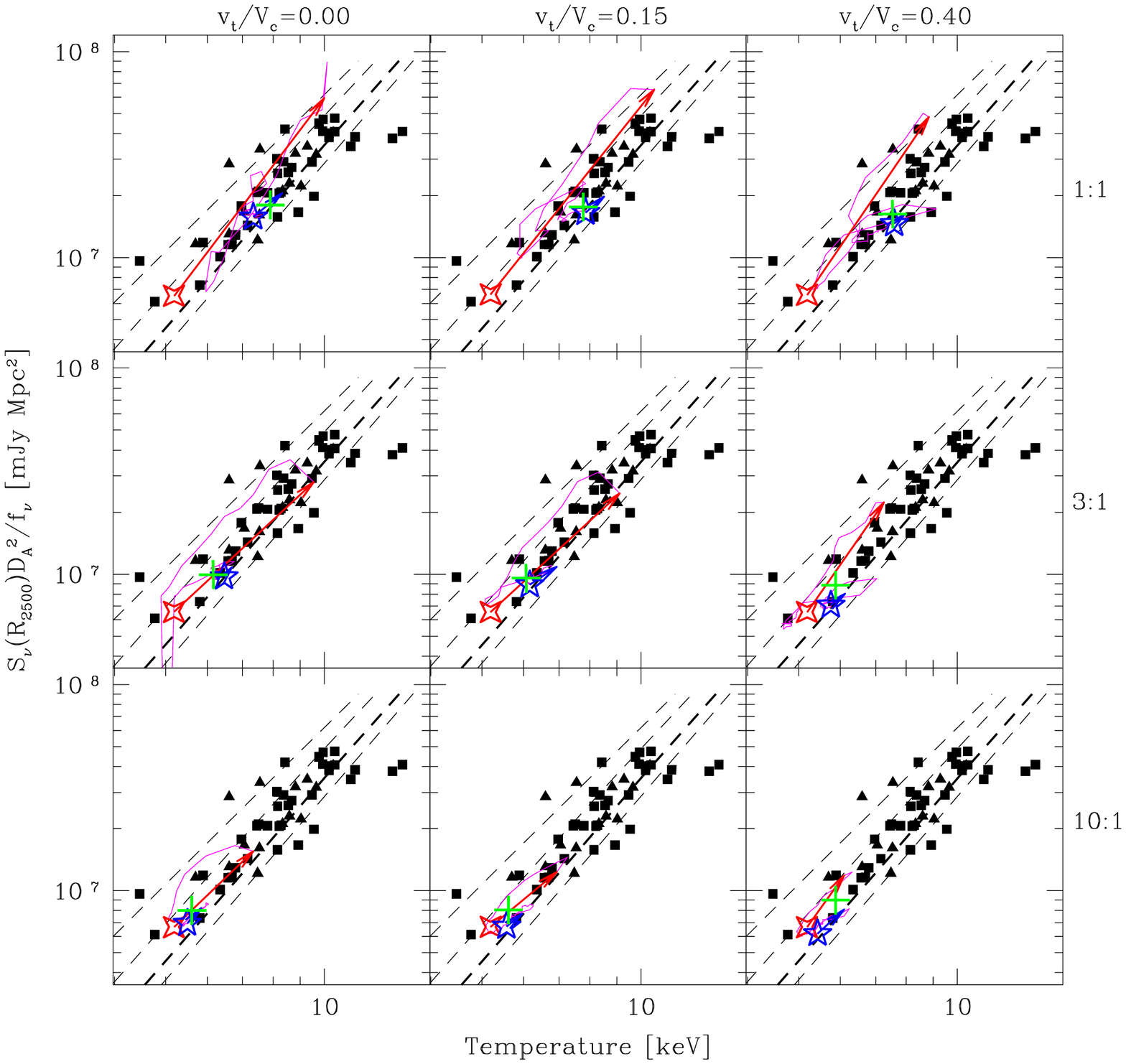}
\caption{Evolution of the integrated SZ Compton parameter ($S_\nu$, integrated within $R_{2500}$) plotted against temperature (integrated within $R_{500}$) for our simulations and compared against observations.  Red four and blue five point stars indicate the states of the system at $t_o$ and \trelax~ respectively.  The red vector indicates the evolution from $t_o$ to \tclosest~ while the blue vector illustrates $3$\Gyr~ of evolution following \trelax.  The magenta line tracks the evolution from \tclosest~ to \trelax~ (generally $4$-$5$\Gyrs) during which the system is visibly disturbed.  Green crosses indicate the remnant state necessary to preserve the observed mass-scaling relations.  Black points illustrate the observed catalogues of \citet[][triangles]{Reeseetal02} and \citet[][squares]{LaRoqueetal06} with temperatures taken from \citet{Horner_thesis}.  Dashed lines are our fiducial analytic entropy injection models ($S_o=10$,$100$,$300$, and $500$\keVcmsq, increasing from left to right).  Black text around the boundary indicates the mass ratio and $v_t/V_c$ depicted by each panel.}
\label{fig-Snu_Tx}
\end{center}
\end{minipage}
\end{figure*}
\begin{figure*}
\begin{minipage}{175mm}
\begin{center}
\leavevmode \hspace*{-1cm} \epsfysize=19cm \epsfbox{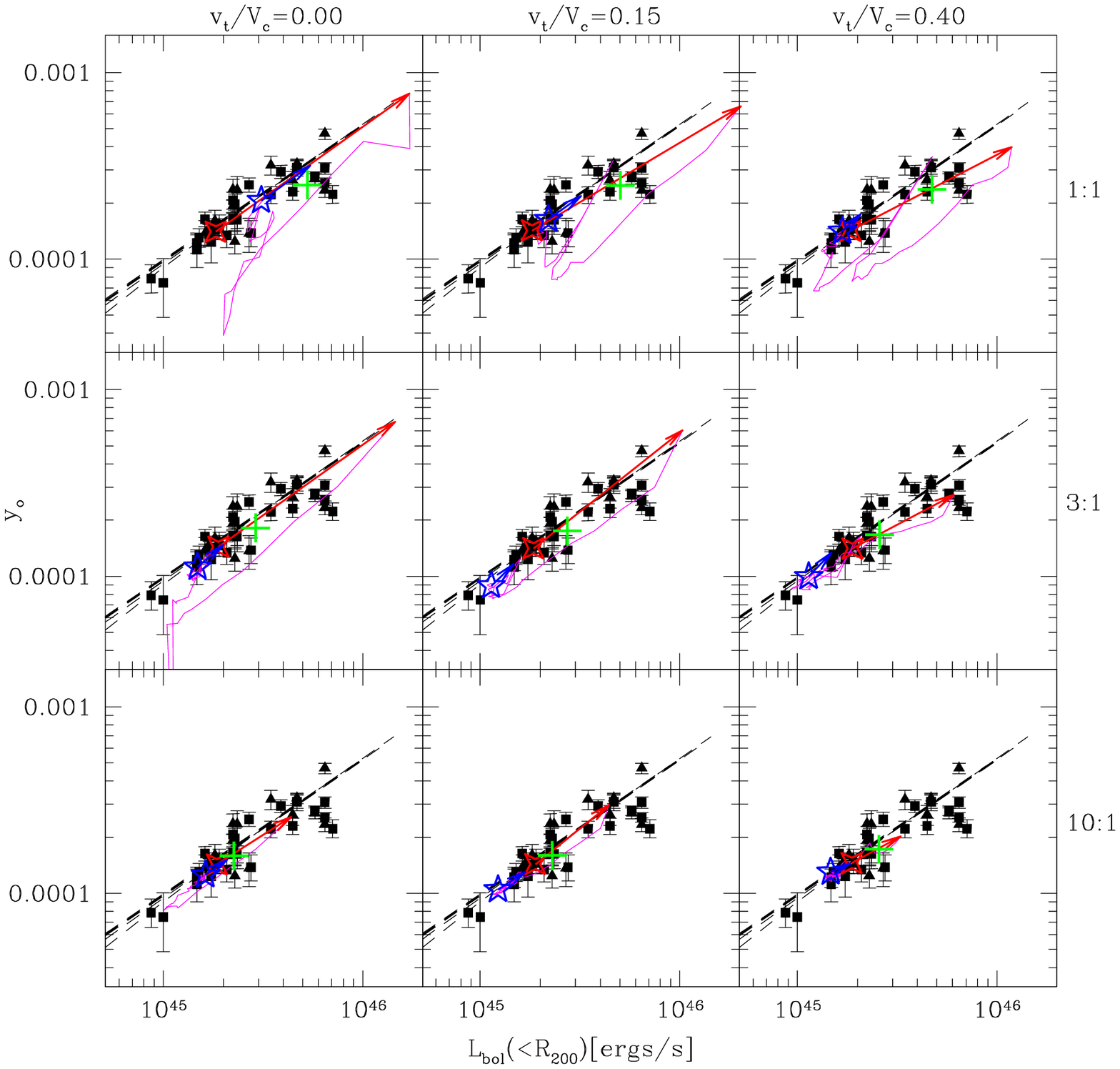}
\caption{Evolution of the SZ Compton parameter ($y_o$) plotted against bolometric luminosity (integrated within $R_{500}$) for our simulations and compared against observations.  Red four and blue five point stars indicate the states of the system at $t_o$ and \trelax~ respectively.  The red vector indicates the evolution from $t_o$ to \tclosest~ while the blue vector illustrates $3$\Gyr~ of evolution following \trelax.  The magenta line tracks the evolution from \tclosest~ to \trelax~ (generally $4$-$5$\Gyrs) during which the system is visibly disturbed.  Green crosses indicate the remnant state necessary to preserve the observed mass-scaling relations.  Black points illustrate the observed catalogues of \citet[][triangles]{Reeseetal02} and \citet[][squares]{LaRoqueetal06} with luminosities taken from \citet{Horner_thesis}.  Dashed lines are our fiducial analytic entropy injection models ($S_o=10$,$100$,$300$, and $500$\keVcmsq, increasing from right to left).  Black text around the boundary indicates the mass ratio and $v_t/V_c$ depicted by each panel.}
\label{fig-yo_Lx}
\end{center}
\end{minipage}
\end{figure*}
\begin{figure*}
\begin{minipage}{175mm}
\begin{center}
\leavevmode \hspace*{-1cm} \epsfysize=19cm \epsfbox{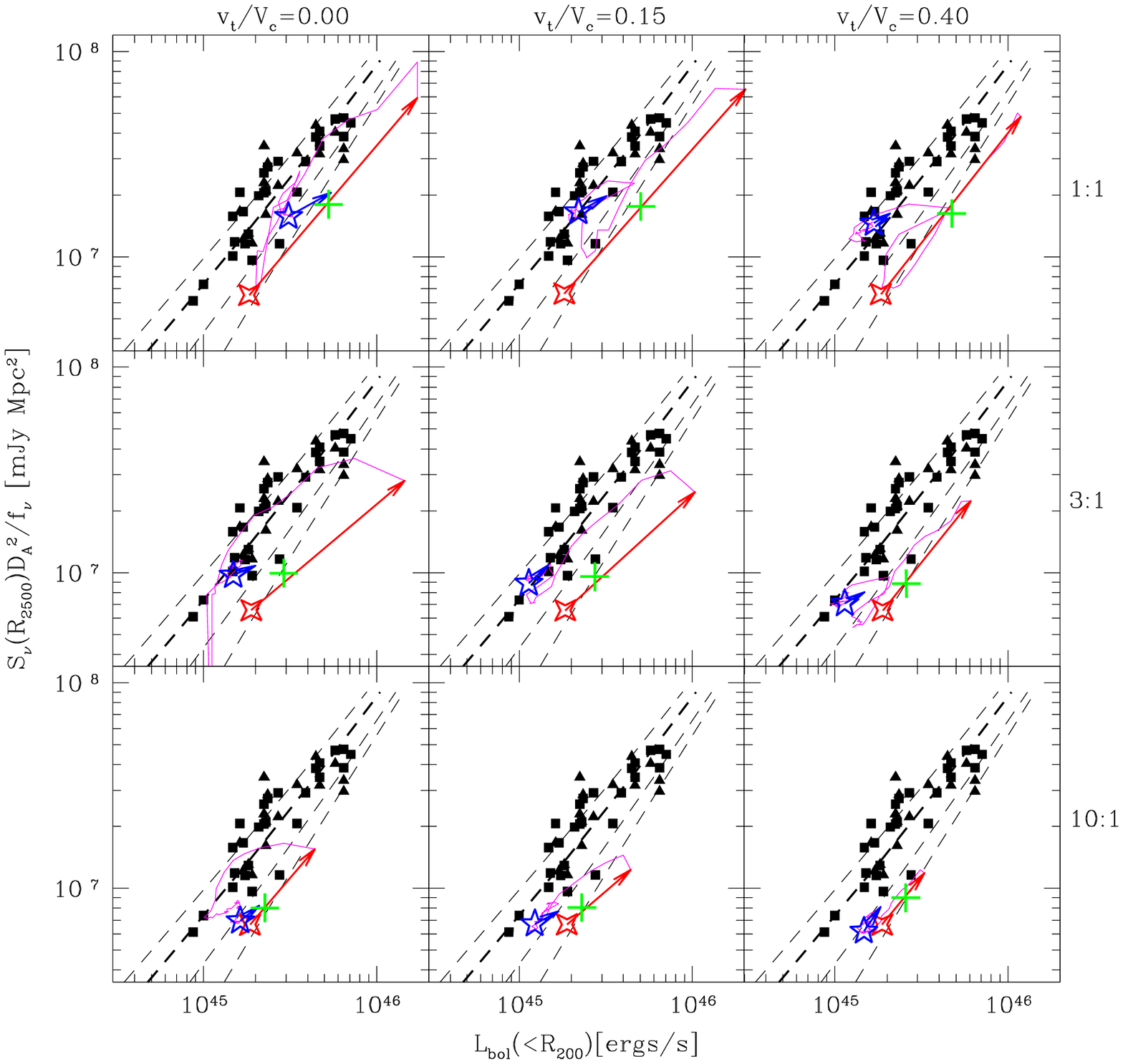}
\caption{Evolution of the integrated SZ Compton parameter ($S_\nu$, integrated within $R_{2500}$) plotted against bolometric luminosity (integrated within $R_{500}$) for our simulations and compared against observations.  Red four and blue five point stars indicate the states of the system at $t_o$ and \trelax~ respectively.  The red vector indicates the evolution from $t_o$ to \tclosest~ while the blue vector illustrates $3$\Gyr~ of evolution following \trelax.  The magenta line tracks the evolution from \tclosest~ to \trelax~ (generally $4$-$5$\Gyrs) during which the system is visibly disturbed.  Green crosses indicate the remnant state necessary to preserve the observed mass-scaling relations.  Black points illustrate the observed catalogues of \citet[][triangles]{Reeseetal02} and \citet[][squares]{LaRoqueetal06} with luminosities taken from \citet{Horner_thesis}.  Dashed lines are our fiducial analytic entropy injection models ($S_o=10$,$100$,$300$, and $500$\keVcmsq, increasing from right to left).  Black text around the boundary indicates the mass ratio and $v_t/V_c$ depicted by each panel.}
\label{fig-Snu_Lx}
\end{center}
\end{minipage}
\end{figure*}

The magnitude of a system's thermal SZ signature as a function of position on the sky (denoted by $\theta$) is typically quantified by the Compton ``y-parameter'' given by
\begin{center}
\begin{equation}\label{eqn-Compton}
y(\theta)=\frac{\sigma_T}{m_ec^2}\int P_e(\vec{r})\,dl
\end{equation}
\end{center}

\noindent where the integral is of the three dimensional pressure $P_e(\vec{r})=n_e(\vec{r})kT(\vec{r})$ at position $\vec{r}$ and is performed along the line of sight through the system.

Compton parameters through the centres of systems ($\theta=0$; denoted $y_o$) are commonly reported.  However, due to the difficulty of correcting observed central decrements for the contamination from negative side lobes introduced by interferometers, observed values are typically determined from fitting an assumed functional form of $y(\theta)$ (convolved with the beam of the instrument) to the observed profile, $y_{obs}(\theta)$.  The SZ profile commonly assumed is that expected for an isothermal $\beta$-model 
\begin{center}
\begin{equation}\label{eqn-Compton_assumed}
y'(\theta)=\frac{y_o}{\left( 1 +(r/r_c)^2 \right)^{(3\beta-1)/2}}
\end{equation}
\end{center}

\noindent with $\beta$ and $r_c$ obtained from X-ray observations of the system.  The normalization of the fit of $y'(\theta)$ to $y_{obs}(\theta)$ yields the observed value of $y_o$ \citep[see][for more detailed accounts of this procedure]{LaRoqueetal06,Reeseetal02}.  

This approach will tend to capture the system's structure outside of $r_c$ and will treat the core as uniform.  Thus, since nearly all systems have $r_c>30$\kpc, variations within $30$kpc~ due to complex AGN processes will tend to be diminished.  In Fig. \ref{fig-yo_t} we demonstrate this fact by plotting the time evolution of $y_o$ (measured in the $z$-projection) for each of our simulations.  The thin blue curve illustrates $y_o$ computed using the $\beta$-model approach presented above, and the thick red curve presents a direct integration of our simulations with the ICM's central ($r<30$\kpc) temperature and density set to it's value at $30$\kpc. The two results are in very good agreement: the $\beta$-model approach often used in observational studies does a very good job of capturing the structure of the system outside of $30$kpc~ over the entire duration of our mergers.

We estimate that the contribution to $y_o$ from material within $30$\kpc~ is $\sim 20$\%.  Hence, large fluctuations due to AGN activity in this region can thus have a significant effect on $y_o$.  Thus, the tendency of the $\beta$-model approach to diminish variations in $y_o$ has both positive and negative consequences.  On the one hand, reducing the sensitivity of $y_o$ to regions whose structure is not well understood and whose evolution is affected by stochastic effects generated by central AGN should lead to much tighter scaling relations.  This is useful for cosmological studies where measures which accurately characterize the total mass of the system are desired.  On the other hand, reducing the sensitivity of this measure to the cluster's most central regions reduces its utility for studying the uncertain structure of cluster cores and the processes active within them.

We have also placed thick dashes on Fig. \ref{fig-yo_t} to represent the expected mass scaling of $y_o$ from its initial values (we assume $y_{o,2}/y_{o,1}=\left( M2/M_1 \right)^{\alpha}$, with $\alpha$ determined to be $0.93$ from our fiducial models since adequate empirical measures are not yet available).  This is performed for the $\beta$-model results which include the central core.  We can see that the expected mass scaling of $y_o$ is recovered in each case to within $30$\%, with a systematic tendancy for $y_o$ to fall below this value with increasing impact parameter.

\citet{Bensonetal04} have pointed out that $y_o$ can be significantly biased by uncertainties in the assumed spatial model of the ICM.  They present an analysis of integrated Compton fluxes which they find to be more robust to uncertainties in the assumed form of $y'(\theta)$.  Integrated SZ fluxes are computed from Eqn. \ref{eqn-Compton} according to

\begin{center}
\begin{equation}\label{eqn-Compton_int}
y_{int}(<\theta)=2\pi \int_{0}^{\theta} y(\theta')\theta'\,d\theta'
\end{equation}
\end{center}

\noindent However, SZ measurements are usually expressed in terms of frequency dependent flux densities which are given by
\begin{center}
\begin{equation}\label{eqn-Compton_flux_density}
S_\nu=j_\nu(x) y_{int}
\end{equation}
\end{center}

\noindent where $x=h\nu/kT$ is a dimensionless frequency and $j_\nu$ is a function which describes the spectral shape of the SZ distortion.  This function is given by
\begin{center}
\begin{equation}\label{eqn-j_nu}
j_\nu(x)=\frac{2(kT_{CMB})^3}{(hc)^{2}}f_\nu(x)
\end{equation}
\end{center}

\noindent where
\begin{center}
\begin{equation}
\label{eqn-f_nu}
f_{\nu}(x)=\frac{x^4 e^x}{(e^x -1)^2} \left( \frac{x}{\tanh(x/2)}-4 \right)
\end{equation}
\end{center}

To facilitate comparisons to observations, which are performed at many different frequencies and for systems at various redshifts, we remove the frequency and distance (redshift) dependencies from our computed flux densities by reporting the quantity $S_\nu D_A^2/f_\nu$.  

In Fig. \ref{fig-Snu_t} we present the evolution of our merging systems' SZ flux integrated within a projected radius of $R_{2500}$ (measured in the $z$-projection).  Once again we illustrate two quantities in each case: the integrated SZ flux computed directly from the simulations and from integrating the best fit of $y'(\theta)$ to $y_{obs}(\theta)$.  We see that integrating the best-fit $\beta$-model does an excellent job of tracking the actual value.  In addition, we have placed thick dashed lines to indicate the expected mass scalings of $S_\nu$ from its initial values (we assume $S_{\nu,2}/S_{\nu,1}=\left(M_2/M_1\right)^\alpha$, with $\alpha$ determined from our fiducial models to be $1.63$, once again, because adequate empirical measures are not yet available).  We see that this SZ measure recovers the mass scaled values far better than $y_o$.  

In all our simulations the evolution of both $S_\nu$ and $y_o$ follows the familiar pattern seen for $L_x(t)$, $T_x(t)$ and $M_t(t)$: a rapid rise that begins shortly before \tclosest, a subsequent drop to values below the initial state, a second (but substantially weaker) peak at \taccrete, a slow decrease until \trelax~ and then a final rise continuing to the end of the simulation.  Radiative and adiabatic cooling (arising from an expansion of the core) results in a significant reduction from the initial value following \taccrete.  As active cooling reestablishes in the remnant cores, the system's integrated SZ signature begins increasing.

For the remainder of this section, we shall examine the scaling relations constructed from these quantities and the X-ray temperature and luminosity.  Observations for comparison to our simulations are still very sparse and statistically incomplete at the masses probed by our simulations, and we present these plots for completeness and to establish theoretical expectations for understanding larger data sets, once they become available.

\subsection{SZ$-T_x$ scaling relations}\label{analysis-SZ_T_scaling}

The scaling behaviours of $y_o$ and $S_\nu$ with $T_x$ are presented in Figs. \ref{fig-yo_Tx} and \ref{fig-Snu_Tx} following the format of our previous scaling relation plots: red four and blue five point stars mark the system's state at $t_o$ (virial crossing) and \trelax~ (when the system appears relaxed and is virialized) respectively, while red vectors illustrate the system's evolution from $t_o$ to \tclosest~ (first pericentric passage) and blue vectors mark $3$\Gyrs~ of evolution following \trelax.  The magenta curve illustrates the evolution of the system while it is visibly disturbed between \tclosest~ and \trelax.  During this time, the system experiences a second transient roughly along the scaling relation and drifts slightly across its scatter.

For observational comparison in our SZ analysis, we use the heterogeneous catalogues of massive systems presented by \citet[][black triangles]{Reeseetal02} and \citet[][black squares]{LaRoqueetal06}.  The observed values of $S_{\nu}$ we present are produced from the values of $y_o$ and the $\beta$-model parameters tabulated by these authors since values of $S_{\nu}$ are not presented in either catalogue.  Given the excellent recovery of $S_{\nu}$ from integration of Eqn. \ref{eqn-Compton_assumed} illustrated above, these values should be an accurate representation of the actual values for these systems.

Examining the $y_o-T_x$ plane first, we note two general trends in our simulations: when the system is most disturbed at \tclosest, it tends to be deflected towards high values of $y_o$ and once relaxed, it evolves very little.  Since $y_o\propto \rho_g T_x$, the first of these observations can be understood as a product of the additional increases in $y_o$ due to higher gas densities generated by compression.  The second can be understood as a consequence of the $\beta$-model treating all systems as constant density isothermal cores within $r_c$.  Even though the system may be relaxing and central pressures increasing as cooling recovers following \trelax, the isothermal $\beta$-model will not catch this behaviour and very little evolution is seen as a result.

The range of central entropies chosen for our fiducial model were selected to encompass the majority of observed systems in the $L_x-T_x$ plane and they generally do so in the other scaling relations we have studied so far.  However, in the $y_o-T_x$ plane (and all the following SZ scaling relations we present) there is a slight bias in the observations towards our high-$S_o$ fiducial models.  This trend has been noted in previous studies as well \citep{McCarthyetal03b}.  Without knowing the degree to which the observed catalogue is statistically representative, it is difficult to comment any further on this.  

In Fig. \ref{fig-Snu_Tx} we present the evolution of our simulations in the $S_{\nu}-T_x$ plane.  Our fiducial models do a better job of embracing the scatter in the observed catalogues in this case, but there is still a slight bias towards the higher-$S_o$ models.  Both the observed scatter and the scatter expected from our fiducial models is considerably less.  

In both planes, we once again see that mergers can displace a system to approximately the $S_o=300$\keVcmsq~ fiducial track, but not to the fiducial tracks of higher-$S_o$ models where many observed systems are located.

\subsection{SZ$-L_x$ scaling relations}\label{analysis-SZ_L_scaling}

In Figs. \ref{fig-yo_Lx} and \ref{fig-Snu_Lx} we present the evolution of our simulations in the $y_o-L_x$ and $S_\nu-L_x$ planes, following the same format as the SZ$-T_x$ scaling relations presented in Section \ref{analysis-SZ_T_scaling}.

Examining the $y_o-L_x$ plane first, we immediately notice the remarkably small variation in the fiducial mass scaling relations with $S_o$, first predicted by \citet{McCarthyetal03}.  Our fiducial models capture the median trend of the observations and we see that in all cases, our simulated systems begin on this predicted scaling relation (red four point star) and by \trelax~ (blue five point star) have returned to it and are evolving along it.

During first pericentric passage (red vector) and while significantly disturbed between \tclosest~ and \trelax~ (magenta curve), the system is over luminous relative to our fiducial models.  At no time do our mergers evolve to states which are significantly under luminous relative to the median trend of the observations.  However, although the scatter in the observations is small, many systems deviate from the fiducial relation by a statistically significant amount.  Those systems which are significantly over luminous relative to our fiducial models can potentially be accounted for through merger activity, but those which are under luminous can not.  Increased pressure from AGN activity could be responsible for the deviant state of such systems.

In Fig. \ref{fig-Snu_Lx} we find that the degeneracy in the SZ$-L_x$ mass scaling relations with $S_o$ is broken when $S_\nu$ is used, with both our fiducial models and the observations exhibiting significantly more scatter.  Our simulations show that while disturbed, the system can swing wildly to states both over luminous with respect to the observations and our fiducial models (during first pericentric passage) and briefly to the lowest luminosity side of the scatter.  Once relaxed however, states represented by our fiducial models with $S_o>300$\keVcmsq~ are not produced.

\section{The X-ray Sunyaev-Zel'dovich proxy}\label{analysis-XSZ}

\begin{figure*}
\begin{minipage}{175mm}
\begin{center}
\leavevmode \epsfysize=8.9cm \epsfbox{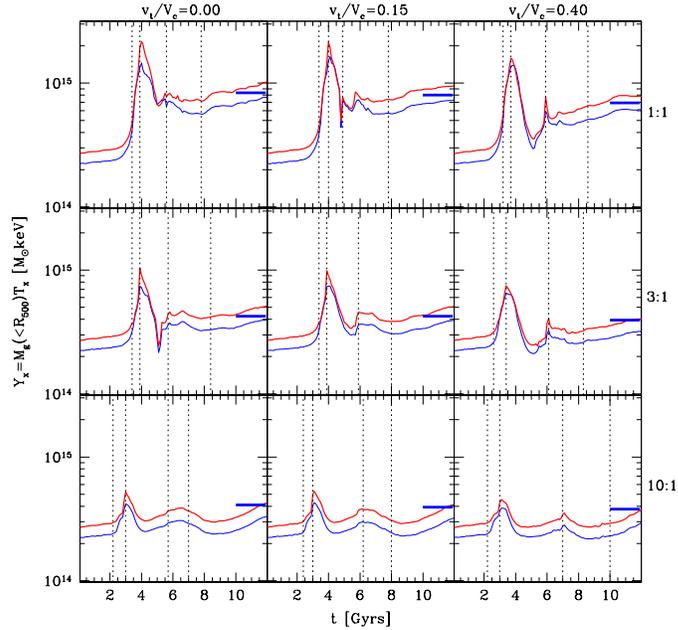}
\caption{The X-ray SZ proxy ($Y_x=M_gT_x$) for our simulations as a function of time.  Thick red curves illustrate values integrated directly from our simulations while thin blue curves are computed according to \citet{Kravtsovetal06}.  Horizontal dashes represent the final values expected from the mass scalings derived from our fiducial models (see Section \ref{analysis-luminosity}).  Vertical dotted lines indicate (from left to right) $t_o$ (the time of virial crossing), \tclosest~ (first pericenter), \taccrete~ (second pericenter) and \trelax~ (the moment our remnants would appear relaxed to a $50$\ks~ \Chandra~ exposure at $z=0.1$).  Black text around the boundary indicates the mass ratio and $v_t/V_c$ depicted by each panel.}
\label{fig-Yx_t}
\end{center}
\end{minipage}
\end{figure*}

Lastly, we have examined the evolution of the X-ray proxy for SZ fluxes introduced by \citet{Kravtsovetal06}.  This proxy is given by
\begin{center}
\begin{equation}
\label{eqn-XSZ}
Y_x(<R)=M_g(<R)T_x
\end{equation}
\end{center}
where $M_g$ is the gas mass within $R$ and $T_x$ is the X-ray temperature measured within $R$, with the central $0.1R_{200}$ excised.  In Fig. \ref{fig-Yx_t} we present the evolution of $Y_x$ integrated within $R_{500}$ as measured directly from our simulations (red) and our isothermal $\beta$-model fits (blue; in the $z$-projection in all cases).  The evolution of this quantity is (as expected) very similar to that of the integrated SZ flux with a similar underestimate of $\sim 15-25$\% by the isothermal $\beta$-model from actual values.  We have once again placed thick dashes to indicate the expected mass scalings (derived from our fiducial models) from initial values.  

\begin{figure*}
\begin{minipage}{175mm}
\begin{center}
\leavevmode \epsfysize=8.9cm \epsfbox{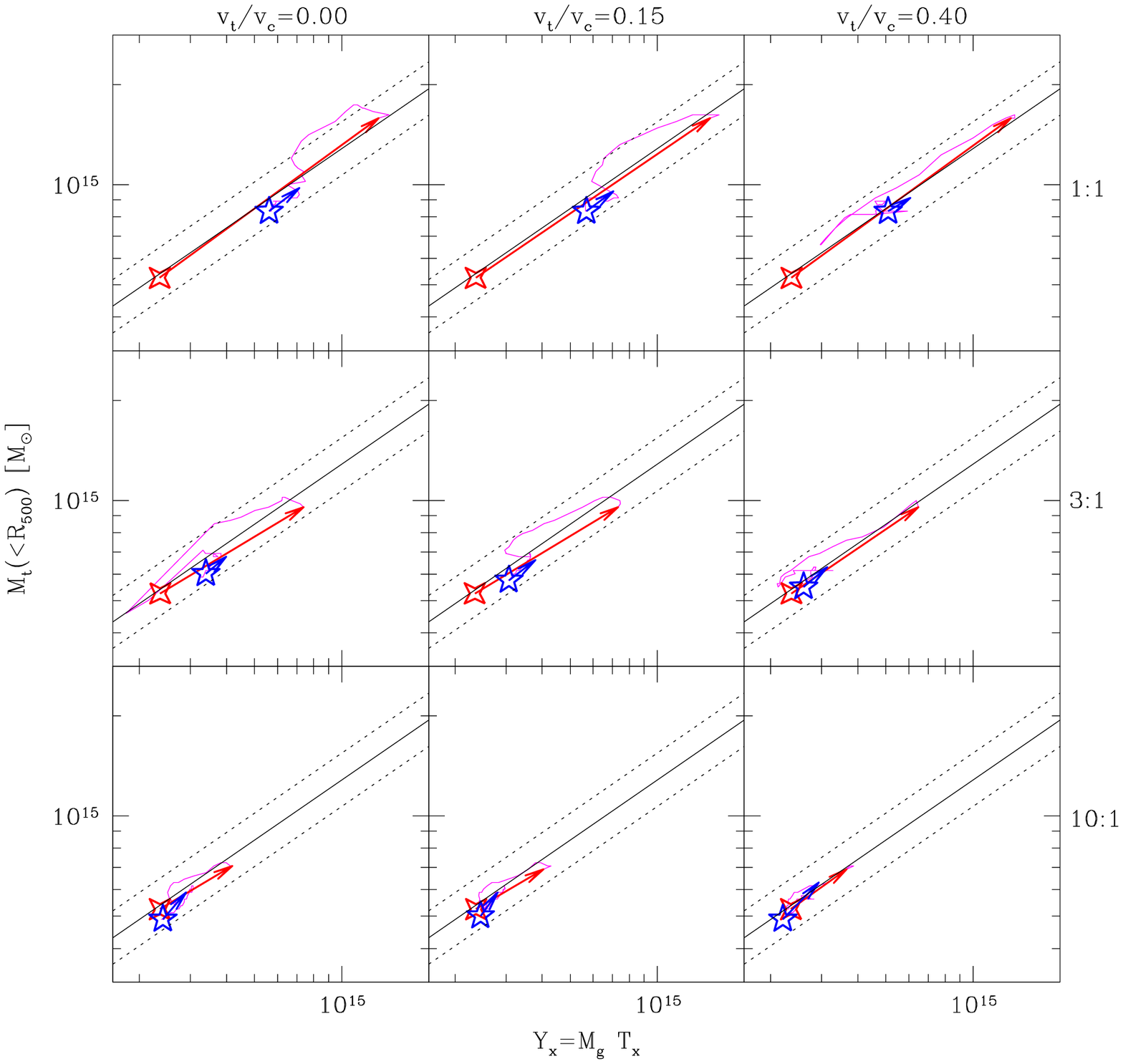}
\caption{Evolution of the X-ray SZ proxy ($Y_x$) introduced by \citet{Kravtsovetal06} plotted against total mass for our simulations (both integrated within $R_{500}$).  Red four and blue five point stars indicate the states of the system at $t_o$ and \trelax~ respectively.  The red vector indicates the evolution from $t_o$ to \tclosest~ while the blue vector illustrates $3$\Gyr~ of evolution following \trelax.  The magenta line tracks the evolution from \tclosest~ to \trelax~ (generally $4$-$5$\Gyrs) during which the system is visibly disturbed.  Green crosses indicate the remnant state necessary to preserve the observed mass-scaling relations.  Black text around the boundary indicates the mass ratio and $v_t/V_c$ depicted by each panel.}
\label{fig-Mt_Yx}
\end{center}
\end{minipage}
\end{figure*}

\subsection{$M_t-Y_x$ relation}\label{analysis-Mt_Yx}

In Fig. \ref{fig-Mt_Yx} we present the scaling of the X-ray SZ proxy with total actual projected mass within $R_{500}$.  Astonishingly, we find that our system remains well confined within the $8$\% scatter bounds of the mean relation determined by \citet{Kravtsovetal06} at all times.  There is a slight tendency for the final remnant to evolve on the low-$Y_x$ side of the relation while visibly disturbed (magenta line) and to evolve to states shifted slightly to the high-$Y_x$ side of the mean following \trelax~ (blue five point star).

We present results for only the $z$ projection in Fig. \ref{fig-Mt_Yx} but have confirmed that the system remains within the $8$\% scatter bounds at all times in all projections.

\section{Discussion}\label{sec-discussion}

Scaling relations are powerful tools for constructing mass functions for cosmological studies.  However, it has been known for some time \citep{Fabianetal94,Markevitch98}, and we have illustrated in this paper, that variations in core properties contribute significantly to the scatter in both X-ray and SZ properties.  This is undesirable for cosmological studies (motivating many authors to exclude the central regions of clusters from their analysis) but affords an opportunity to understand the processes which shape cluster cores.  Such processes include, but are not restricted to, the effects of mergers, AGN and various other processes (such as preheating) which may have contributed to modifying the central entropy (and hence, morphology) of cluster cores.  It is important to keep this dual role of cluster scaling relations in mind when considering global cluster properties: both core corrected and uncorrected values are of great utility.

Keeping the cores in our analysis, we have examined scaling relations between luminosity, temperature, mass and SZ properties.  In most cases, we find that relaxed states which cover the full dispersion in the scaling relations can only be achieved through extremely rare equal-mass, off-axis mergers.  Somewhat more common 3:1 mergers are rarely able to generate systems (neither transiently nor as relaxed remnants) which account for the dispersion in the observed scaling relations.   In order for mergers to move the system the rest of the way across the relation, an additional merger of comparable extent would thus be required.  To have a significant effect on the statistics of the population this would have to occur regularly, requiring the compounding of infrequent events.  In short, mergers between typical relaxed compact cool core systems can not account for the dispersion in observed scaling relations.  

Ultimately, detailed theoretical studies of the scatter in scaling relations must be conducted with cosmological hydrodynamic simulations.  To date, \citet{Kayetal06} and \citet{OHaraetal05} are the only two cosmological N-body studies which have specifically examined the scatter in scaling relations.  Both of these studies support the results of our analysis: \citet{OHaraetal05} find that their simulations clearly fail to produce enough scatter to account for observations and \citet{Kayetal06} find that nearly all of their clusters, regular or irregular, have cool cores.  Since approximately half the scatter in observed scaling relations is due to NCC systems with warm cores \citep[see][]{McCarthyetal07b}, we infer that their simulations fail to reproduce the observed scatter as well.  Furthermore, it is also worth noting that the study we have presented in this paper will be useful as a tool for interpreting the effects of individual events for future scaling relation studies conducted with cosmological simulations.

We have initialized our systems to represent common compact cool core systems observed by \Chandra~ and to represent a significant range of orbital properties found in cosmological dark matter simulations.  As a consequence, our systems should represent a significant range of typical cluster merger events.  A natural question to ask however, is whether systems with initially higher central entropies could be driven by mergers to states which account for the dispersion in the scaling relations.

Such a scenario is effectively a multiple heating event scenario in which a system's central entropy becomes elevated (either by preheating, a previous merger event or post-collapse AGN activity) and then additionally heated, before it is able to relax, to states which single mergers between relaxed systems can not generate.  This is appealing on the grounds that mergers involving heated cores will generate lower densities and stimulate less radiative cooling, tipping the balance between heating and cooling towards heating.  Furthermore, previously heated cores will have lower densities, increasing the efficiency of entropy production \citep{Voitetal03} from shocks.

One appealing mechanism (which we shall examine in a future study) of generating frequent multiple heating events could be the stimulation of AGN activity through mergers.  There are several stages during the evolution of our mergers when gaseous material is driven towards the cores of our systems.  These times generally occur during the interactions of the merging cores (at \tclosest~ or \taccrete) or during the accretion of the streams generated from the disruption of the secondary's core.  This may provide a natural means of generating multiple heating events in close succession (and thus reproducing the full dispersion of the scaling relations) within the framework of hierarchical clustering.

Alternatively, \citet{Burnsetal06} have proposed a plausible means by which cluster mergers (albeit between unrelaxed systems at high redshift) could account for the observed dispersion in cluster properties.  Using a set of cosmological adaptive mesh refinement (AMR) simulations, they argue that early major mergers can prevent the establishment of cold cores, leading to the production of systems with high central entropies.  Although this result is preliminary and has not yet been reproduced by other authors (particularly those utilizing SPH algorithms), it is an interesting possibility worth considering.

Impending observational progress should contribute significantly towards resolving these issues.  We have shown for example that using the isothermal $\beta$-model to measure masses not only leads to a downward bias of $\sim 30$\%, but likely also contributes significantly to the observed scatter in mass scaling relations.  However, it has been demonstrated \citep[so far only on limited samples;\eg][]{Vikhlininetal06} that significant improvements in X-ray mass estimates are feasible with high quality X-ray observations using modern reduction techniques.  Once applied to a large representative sample of clusters, interpretation of observed mass-scaling relations will be greatly simplified.

Furthermore, improvements in available SZ catalogues will be extremely valuable, particularly when combined with X-ray and weak lensing maps \citep[see][for a discussion of the benefits such a cross correlation will afford]{Mahdavietal06}.

\section{Summary}\label{sec-summary}
Building on the results of previous authors \citep[most notably,][]{RickerandSarazin01}, we have identified a generic evolutionary sequence for X-ray and SZ observables during cluster mergers.  In all the cases we have studied (which we believe represent a significant range of cluster merger initial conditions), these properties (specifically $L_x$, $T_x$, $M_t$, $y_o$ and $S_\nu$) exhibit large increases during core-core interactions at \tclosest~ and \taccrete~ with an intervening drop from initial values due to the expansion of the primary system in response to the initial core-core interaction.  Our systems then experience a subsequent rise in all quantities following \trelax~ due to the reestablishment of efficient radiative cooling in the remnent cores and the reaccretion of displaced material.  This increase lasts until the conclusion of our simulations.

Because of this common evolution in the global properties we have studied, merger systems tend to evolve in a generic way in scaling relations constructed from them: there is an initial transient which evolves roughly along the mass scaling relations, followed by a drift across the relation until \trelax, after which the system quiescently evolves along the mass scaling relation.

Our study indicates that typical cluster mergers fail to generate states which can account for the scatter of observed scaling relations.  They can push relaxed clusters to the median position of the observed scaling relations, but not past it (not even transiently).

Several other, more specific findings of our study, include:

\begin{itemize}
\item The luminosity integrated within $R_{500}(t)$ recovers the observed mass scaling relations only in the head-on cases.  As mergers are moved off-axis, the remnant's luminosity becomes steadily less than expected from the observed mass scaling (discrepancies are $16-50$\%).  This discrepancy is still present but is reduced to levels of $3-28$\% if the central regions ($r<0.1R_{200}$) are excised.

\item We have studied the evolving temperatures of our mergers using 3 temperature measures: emission weighted temperatures ($T_{ew}$),  the ``spectroscopic-like'' average presented by \citet{Mazzottaetal04} ($T_{sl}$) and isothermal fits to integrated spectra ($T_{spec}$).  We find that $T_{ew}$ is higher than the other two measures at all times and that $T_{spec}$ produces results which agree with $T_{sl}$ to within $5$\% at nearly all times.  We thus confirm the utility of this method for generating globally averaged temperatures from massive simulated clusters.

\item $T_{ew}$ yields remnant temperatures significantly in excess of what is expected from the observed mass scaling relation while the other measures recover the observed mass scaling result in all cases to within $15$\%.  This should lead to significantly less scatter in theoretical $L_x-T_x$ relations computed with $T_{sl}$ or $T_{spec}$.

\item The temperature spikes following first and second pericentric passage (at \tclosest\ and \taccrete\ respectively) are $15-40$\% larger for the $T_{ew}$ measure than the other two methods.  This suggests that the estimates of the effect of merger-driven temperaure spikes on mass functions determined from temperature scaling relations utilizing $T_{ew}$ \citep[\eg][]{Randalletal02} have been exagerated.

\item As a consequence of the mass scaling behaviours of $L_x$ and $T_x$, head-on mergers ultimately move systems along the observed $L_x-T_x$ scaling relation while off-axis mergers scatter systems towards lower luminosities.

\item Isothermal $\beta$-models track the evolution of the actual mass of the system before the interaction and while appearing relaxed afterwards, but with a systematic underestimate of $25-40$\% supporting the findings of several previous studies.  

\item The isothermal $\beta$-model does a very good job of reproducing the actual results for the SZ observables $y_o$ and $S_\nu (<R_{2500})$.  Theoretical mass scaling relations are preserved much better by the integrated measure.

\item Remarkably little scatter is expected in the $y_o-L_x$ plane \citep{McCarthyetal03,McCarthyetal03b}.  Mergers push systems only to the high luminosity side of the scatter and the system relaxes back to the median relation afterwards.  These results make it difficult to use mergers to account for under luminous systems in this plane.

\item Regardless of projection effects, the SZ proxy of \citet{Kravtsovetal06} obeys their reported mass scaling to within $8$\% over the entire duration of all our merger events.
\end{itemize}

We conclude that the full dispersion in scaling relations can not be produced from simple two body mergers between common relaxed compact cool core systems.  However, some displacement across each plane does occur and this is driven primarily by changes in the structure of a cluster's core.  Thus, the outer (\ie~ $r>0.1R_{200}$) structure of clusters are preserved following merger events.  This is satisfying in light of recent observations suggesting that the outer structure of clusters exhibit a universal structure.

Mergers certainly alter the structure of compact cool cores (although not sufficiently to account for the observed range of system properties) however, and we shall explore the details of these effects and the processes driving them in the next paper of this series.

\section*{Acknowledgements}
We are grateful to J. Cohn for providing us with our merger rate statistics (and to Martin White for making his simulations available to do so).  We would also like to thank Michael Balogh and Neal Katz for stimulating discussions and insightful comments.  IGM acknowledges support from a NSERC Postdoctoral Fellowship and a PPARC rolling grant for extragalactic astronomy and cosmology at the University of Durham.  AB acknowledges support from NSERC through the Discovery Grant program.  MF acknowledges support from NSERC, NASA, and NSF.

\label{lastpage}
\end{document}